\documentclass[sigconf]{acmart}
\AtBeginDocument{%
  \providecommand\BibTeX{{%
    \normalfont B\kern-0.5em{\scshape i\kern-0.25em b}\kern-0.8em\TeX}}}
    
\acmConference[ESEC/FSE 2022]{The 30th ACM Joint European Software Engineering Conference and Symposium on the Foundations of Software Engineering}{14 - 18 November, 2022}{Singapore}

\usepackage[turnoff]{notes}

\usepackage{hyperref}
\usepackage{boldline}
\usepackage{color,colortbl}
\usepackage{bigstrut}
\usepackage[ruled]{algorithm2e}
\usepackage{amsmath,amsfonts}
\usepackage{amsmath}
\usepackage{array}
\usepackage{algorithmic}
\usepackage{balance}
\usepackage{blindtext}
\usepackage{booktabs}
\usepackage{caption}
\usepackage{hyperref}
\usepackage[noabbrev]{cleveref}
\usepackage{comment}
\usepackage{enumitem}
\usepackage{eqparbox}
\usepackage{fancybox}
\usepackage{fancyvrb}
\usepackage{framed}
\usepackage{graphicx}
\usepackage{ifthen}
\usepackage{listings}
\usepackage{mathrsfs}
\usepackage{mdwmath}
\usepackage{mdwtab}
\usepackage{multirow}
\usepackage{pifont}
\usepackage{textcomp}
\usepackage{url}
\usepackage{xspace}
\usepackage[normalem]{ulem}
\usepackage[framemethod=TikZ]{mdframed}
\usepackage{caption}
\usepackage{subcaption}
\usepackage{tabularx}
\usepackage{titlesec}
\usepackage{tcolorbox}
\tcbuselibrary{listings,skins}
\usepackage{mathtools}
\usepackage{blindtext}
\usepackage{lstautogobble}
\DeclareGraphicsExtensions{.pdf,.jpeg,.png}
\graphicspath{{figures/}}

\newcommand{\rom}[1]{\uppercase\expandafter{\romannumeral #1\relax}}

\newcommand{\eg}{\hbox{\emph{e.g.,}}\xspace}
\newcommand{\ie}{\hbox{\emph{i.e.,}}\xspace}

\newcommand{\wrt}{\hbox{\emph{w.r.t.}}\xspace}

\newlength\Linewidth
\def\findlength{\setlength\Linewidth\linewidth
\addtolength\Linewidth{-4\fboxrule}
\addtolength\Linewidth{-3\fboxsep}
}
\newenvironment{examplebox}{\par\begingroup
  \setlength{\fboxsep}{3pt}\findlength
  \setbox0=\vbox\bgroup\noindent
  \hsize=0.90\linewidth
  \begin{minipage}{0.90\linewidth}\normalsize}
    {\end{minipage}\egroup
    \textcolor{gray20}{\fboxsep1.2pt\fbox
     {\fboxsep5pt\colorbox{gray05}{\normalcolor\box0}}}
    \endgroup\par\noindent
    \normalcolor\ignorespacesafterend}

\newcounter{RQACounter}

\usetikzlibrary{shadows}
\usepackage{graphics}
\newmdenv[
    tikzsetting= {fill=blueish},
    skipabove=0.33em,
    skipbelow=0.33em,
    linewidth=1pt,
    innerleftmargin=4pt,
    innerrightmargin=4pt,
    innertopmargin=2pt,
    innerbottommargin=2pt,
    linecolor=gray95,
    roundcorner=2pt, 
    shadow=true,
    shadowsize=4pt,
    shadowcolor=gray95
]{questionbox}

\newmdenv[
    tikzsetting= {fill=greenish},
    skipabove=0.33em,
    skipbelow=0.33em,
    linewidth=1pt,
    innerleftmargin=4pt,
    innerrightmargin=4pt,
    innertopmargin=2pt,
    innerbottommargin=2pt,
    linecolor=gray95,
    roundcorner=2pt, 
    shadow=true,
    shadowsize=4pt,
    shadowcolor=gray95
]{answerbox}

\newmdenv[
    skipabove=0.33em,
    skipbelow=0.33em,
    innerleftmargin=4pt,
    innerrightmargin=4pt,
    innertopmargin=2pt,
    innerbottommargin=2pt,
]{lessonbox}

\usepackage{tikz}
\newenvironment{lesson}
{
    \begin{lessonbox}
}
{\end{lessonbox}}

\newenvironment{result}
{\begin{answerbox}}
{\end{answerbox}}

\newenvironment{question}
{\begin{questionbox}}
{\end{questionbox}}

\definecolor{javared}{rgb}{0.6,0,0} 
\definecolor{javagreen}{rgb}{0.25,0.5,0.35} 
\definecolor{javapurple}{rgb}{0.5,0,0.35} 
\definecolor{javadocblue}{rgb}{0.25,0.35,0.75} 

\lstdefinestyle{basejava}{
  language=java,
  showstringspaces=false,
  basicstyle=\small\ttfamily,
  keywordstyle=\bfseries\color{javapurple},
  commentstyle=\itshape\blue,
  identifierstyle=\blue,
  frame=none,
  backgroundcolor=\color{white},
}

\lstdefinestyle{CustomJava}{
  belowcaptionskip=\baselineskip,
  breaklines=true,
  xleftmargin=\parindent,
  language=java,
  showstringspaces=false,
  basicstyle=\scriptsize\ttfamily,
  keywordstyle=\bfseries\color{javapurple},
  commentstyle=\itshape\blue,
  identifierstyle=\blue,
  belowskip=1pt,
  numbers=left,
  gobble=0
}

\lstdefinestyle{CustomJavaWoNumbers}{
  belowcaptionskip=0.5\baselineskip,
  breaklines=true,
  xleftmargin=\parindent,
  language=java,
  showstringspaces=false,
  basicstyle=\scriptsize\ttfamily,
  keywordstyle=\bfseries\color{javapurple},
  commentstyle=\itshape\blue,
  identifierstyle=\blue,
  belowskip=0.5pt,
  numbers=none,
  gobble=0
}

\lstdefinestyle{codit}{
  belowcaptionskip=\baselineskip,
  breaklines=true,
  language=java,
  showstringspaces=false,
  basicstyle=\scriptsize\ttfamily,
  keywordstyle=\bfseries\color{javapurple},
  commentstyle=\itshape\blue,
  identifierstyle=\blue,
}

\DeclareMathOperator*{\argmin}{arg\,min}

\newcommand\red[1]{\textcolor[rgb]{1.00,0.00,0.00}{#1}}

\newcommand\blue[1]{\textcolor[rgb]{0.00,0.00,1.00}{{#1}}}

\definecolor{blueish}{RGB}{250, 250, 255}
\definecolor{greenish}{RGB}{250, 255, 250}
\definecolor{redish}{RGB}{255, 200, 200}

\definecolor{highlight}{RGB}{175, 255, 100}
\definecolor{gray01}{gray}{.98}
\definecolor{gray05}{gray}{0.95}
\definecolor{gray08}{gray}{0.92}
\definecolor{gray10}{gray}{0.90}
\definecolor{gray12}{gray}{0.88}
\definecolor{gray15}{gray}{0.85}
\definecolor{gray18}{gray}{0.82}
\definecolor{gray20}{gray}{0.80}
\definecolor{gray25}{gray}{0.75}
\definecolor{gray30}{gray}{0.70}
\definecolor{gray35}{gray}{0.65}
\definecolor{gray40}{gray}{0.60}
\definecolor{gray45}{gray}{0.55}
\definecolor{gray50}{gray}{0.50}
\definecolor{gray55}{gray}{0.45}
\definecolor{gray60}{gray}{0.40}
\definecolor{gray65}{gray}{0.35}
\definecolor{gray70}{gray}{0.30}
\definecolor{gray75}{gray}{0.25}
\definecolor{gray80}{gray}{0.20}
\definecolor{gray85}{gray}{0.15}
\definecolor{gray90}{gray}{0.10}
\definecolor{gray95}{gray}{0.05}

\xspace%

\newcommand{\Comment}[1]{}

\newcommand{\tool}{\textsc{NatGen}\xspace}

\newcommand{\linecode}[1]{\lstinline[escapechar=@,basicstyle=\ttfamily]{#1}~}

\renewcommand{\cite}[1]{\citep{#1}}

\newtcbox{\inlinebox}[1][]{enhanced,
 box align=base,
 nobeforeafter,
 colback=blueish,
 size=small,
 left=0pt,
 right=0pt,
 boxsep=2pt,
 #1}

\newcommand{\highlight}[1]{%
{\footnotesize%
\inlinebox{#1}%
}}

\newcommand{\RQrepeat}[2]{%
    \noindent\textbf{RQ#1.~#2}
}

\newcommand{\RQ}[2]{%
    \begin{question}{
        \medskip
        \refstepcounter{RQACounter} \label{rq-#1}
        \noindent\textbf{{RQ\arabic{RQACounter}.~#2}}
    }\end{question}
}

\newcommand{\RS}[2]{%
    \begin{result}
        \textbf{Result {\ref{rq-#1}}:~}{\emph {#2}}%
    \end{result}
}

\renewcommand{\cref}[1]{\Cref{#1}}

\newcommand{\rqa}{Does ``Naturalization'' help to improve code generation?} 

\newcommand{\rqb}{How do different components in \tool contribute to code generation?} 

\newcommand{\rqc}{How effective is \tool when fine-tuned for different generative tasks in source code?}
\newcommand{\rqd}{How well does \tool's pre-training help in tasks where labelled data is scarce?}

\newcommand{\errorcode}[1]{
\colorbox{redish}{{\tt \bf {\small #1}}}
}

\begin{document}

\title{\tool : Generative pre-training by ``Naturalizing'' source code}

\author{Saikat Chakraborty}
\affiliation{%
  \institution{Columbia University}
  \city{New York}
  \state{NY}
  \country{USA}
}
\email{saikatc@cs.columbia.edu}

\author{Toufique Ahmed}
\affiliation{%
  \institution{University of California, Davis}
  \city{Davis}
  \state{CA}
  \country{USA}
 }
\email{tfahmed@ucdavis.edu}

\author{Yangruibo Ding}
\affiliation{%
  \institution{Columbia University}
  \city{New York}
  \state{NY}
  \country{USA}
}
\email{yrbding@cs.columbia.edu}

\author{Premkumar Devanbu}
\affiliation{%
  \institution{University of California, Davis}
  \city{Davis}
  \state{CA}
  \country{USA}
 }
\email{ptdevanbu@ucdavis.edu}

\author{Baishakhi Ray}
\affiliation{%
  \institution{Columbia University}
  \city{New York}
  \state{NY}
  \country{USA}
}
\email{rayb@cs.columbia.edu}

\renewcommand{\shortauthors}{Chakraborty et al.}

%
%



\begin{abstract}
Pre-trained Generative Language models (\eg PLBART, CodeT5, SPT-Code) for source code yielded strong results on several tasks in the past few years, including code generation and translation. These models have adopted varying pre-training objectives to learn statistics of code construction from very large-scale corpora in a self-supervised fashion; the success of pre-trained models largely hinges on these pre-training objectives. This paper proposes a new pre-training objective, ``Naturalizing'' of source code, exploiting code's bimodal, dual-channel (formal \& natural channels) nature. Unlike natural language, code's bimodal, dual-channel nature allows us to generate semantically equivalent code at scale.  
We introduce six classes of semantic preserving transformations to introduce un-natural forms of code, and then force our model to produce more natural original programs written by developers. 
Learning to generate equivalent, but more natural code, at scale, over large corpora of open-source code, without explicit manual supervision, helps the model learn to both ingest \& generate code. 
We fine-tune our model in three generative Software Engineering tasks: code generation, code translation, and code refinement with limited human-curated labeled data and achieve state-of-the-art performance rivaling CodeT5. 
We show that our pre-trained model is especially competitive at zero-shot and few-shot learning, and better at learning code properties (e.g., syntax, data flow).
\end{abstract}
\maketitle
\section{Introduction}
\label{sec:intro}
Statistical models of the ``naturalness" of code~\cite{hindle2012naturalness} have proven useful for a range of  Software Engineering tasks~\cite{allamanis2018survey, pradel2021neural}, including code generation~\cite{amodio2017neural}, repair~\cite{chakraborty2020codit, tufano2019empirical},  summarization~\cite{liu2020retrieval},  retrieval~\cite{parvez2021retrieval}, and clone detection~\cite{white2016deep, ding2021contrastive}. 
The earlier work in this area trained models directly on tasks, including the early work on  type recovery~\cite{hellendoorn2017deep, allamanis2020typilus}, de-obfuscation~\cite{raychev2014code,vasilescu2017recovering},  repair~\cite{gupta2017deepfix}, and summarization~\cite{iyer2016summarizing, ahmad2020summarization}. 
Training on-task requires a lot of labeled data.
While labeled data is abundant for tasks like code completion (where the corpus inherently provides supervision), other tasks like code generation, translation, summarization, repair, etc., require well-curated, high-quality data.
Simply grabbing data from Github might yield poor-quality~\cite{gros2020code}, highly-duplicated data~\cite{allamanis2019adverse}. 
With increasing model capacity (hundreds of millions, even billions of parameters, are pretty common; larger models tend to perform better~\cite{chen2021evaluating, wang2021codet5}), this unacceptable disparity between vast model capacity and the limited availability of well-curated, high-quality, labeled data has increased and will likely worsen. 
 
 This shortage of high-quality labeled data for on-task training is not unique to Software Engineering (SE), although it is
 complicated here by the increased, specialized skill required
 for labeling SE data. 
To address the issue of training large models in the presence of data scarcity, such models are often pre-trained on some generic tasks, which relate to actual downstream tasks.
 For example, consider two SE tasks: code generation and code translation. Both  tasks require ML models to learn how to generate natural, syntactically, and semantically correct code. This commonality across tasks motivates a quest for better pre-trained models, using a self- (or un-) supervised task 
 which transfers well to other downstream tasks.
Such pre-trained models can also learn a generic representation of the input data, which, in turn, transfers to diverse downstream tasks. 

A popular approach for dealing with this problem involves derivatives of BERT style models~\cite{devlin2018bert}, \eg CodeBERT~\cite{feng2020codebert}, GraphCodeBERT~\cite{guo2020graphcodebert}, etc. These models are good at capturing generic code representations. For code generation tasks, GPT-3 or BART-style models (\emph{e.g.,} Codex, CodeT5,  PL\-BART, SPTCode, etc.~\cite{chen2021evaluating, ahmad2021unified, wang2021codet5, niu2022spt}) are popular. The important insight here
is that independent of final tasks, when \emph{very} high capacity models are trained with huge code corpora to learn simple, self-supervised, ``busy work'', they still \emph{learn general syntactic and semantic constraints of writing code}. Different approaches adopt different techniques to train the model to write code. For instance, GPT-style models (\eg Codex) learn to generate code sequentially, mimicking the left-to-right language model. CodeT5 masks out some tokens and asks the model to generate {\em only} those masked tokens. On the other hand,  PLBART and SPT-Code present the model with erroneous code  (with deleted or masked tokens) and ask the model to generate the corrected, complete code. The  models' ability to generate code depends mainly on the pre-training objective that the model is optimized for. 


We propose a novel pre-training task: we ask the model to ``naturalize" code, \ie take ``weird", synthetic code as input and output semantic equivalent, ``natural" code that a human developer would have written. This is a very demanding pre-training task---the model has to learn both code naturalness \underline{\emph{and}} code semantics. We were inspired by noting the work of human Editors (of books, journals, newspapers): they digest imperfectly written but mostly correct text, understand the intent, and then produce more perfect text with pretty much the same meaning. Editing is \underline{\emph{hard}}: a skilled Editor has to have very high levels of language comprehension, to understand given, potentially badly-written text, \emph{and then} deploy very high-level writing skills to generate well-formed text. If Editing could be used as an at-scale pre-training task, the learned model would presumably have excellent language comprehension and also generate excellent text. However, it's not 
obvious how to generate at-scale training data for this ``Editing" task, say, for English. 

\lstset{escapechar=@,style=CustomJavaWoNumbers}
\begin{figure}[!htb]
    \centering
{
\begin{tabular}{p{0.48\linewidth}|p{0.48\linewidth}}
\begin{center}
    \footnotesize \underline{a. Natural Code}
\end{center}
\begin{lstlisting}
Scanner sc = new Scanner(...);
while (sc.hasNext()) {
   String ln = sc.next();
   ...
}
...
\end{lstlisting}    &  
\begin{center}
    \footnotesize  \underline{b. Un-natural code}
\end{center}
\begin{lstlisting}
Scanner sc = new Scanner(...);
@{\bf \red{for ( ;}}@ sc.hasNext() @{\bf \red{;}} @) {
   String ln = sc.next();
   ...
}
...
\end{lstlisting}\vspace{1mm}
\\
\end{tabular}
}
 \caption{Example of a natural code fragment written by developers and its `un-naturally' transformed counterpart. If the {\tt initialization} and {\tt update} part of the {\tt for} loop were to left empty, developers would write the {\tt while} loop.}
\label{fig:code_for_into}
\end{figure}

But our concern here is code, not natural language. We start with the argument that, because of the bimodal, dual-channel nature of code~\cite{casalnuovo2020theory}, it is indeed possible to generate at-scale training data for the Editing task (a.k.a. refactoring in Software Engineering  terminology). Code has a formal channel, with well-defined semantics; because of this, it's possible to transform code into endless forms, all \emph{meaning-equivalent}. Essentially, we can deploy a set of meaning preserving transformations to \emph{rewrite} existing code from widely-used GitHub projects (which presumably have good-quality code that has passed human code review). These rewrites, (\emph{e.g.,} \Cref{fig:code_for_into}), preserve meaning but will make the code into an artificial, often unnatural form\footnote{Studies, with  human-subjects~\cite{casalnuovo2020does,casalnuovo2020programmers} suggest that humans find such rewritten but semantically identical forms harder to read and understand.}. 


Nevertheless, we now have a matched pair of two semantically equivalent forms of code: a ``de-naturalized" form and the original ``natural" form. Furthermore, we can produce these pairs at-scale, and then pre-train on a code ``Naturalization" task. By analogy with human Editors as described above, such pre-training forces the model to learn two hard things: 1) capture the meaning of the input code, and 2) generate an output that more closely resembles human-written code. We hypothesize that the resulting model will both learn better meaning representations, \emph{and also} generate better code.

To this end, we pre-trained our \tool model, using ``Code Naturalizing'' task. \tool is based on a transformer-based sequence-to-sequence model, and learns 
to ``naturalize" artificially generated ``de-naturalized" code back into the form originally written by developers. 
We {\em emphasize} that \tool learns to generate the whole code; this learned skill  transfers to 
downstream fine-tuning tasks that require code generation. 
We show that our pre-training objective helps model generate more natural code (complete code, with high syntactic and semantic similarity with the original human-written code). With proper fine-tuning, \tool achieves state-of-the-art performance in various downstream fine-tuning tasks, including code generation, code translation, bug fix, that demand code generation. We also show that \tool is specially effective when labelled data is scarce. 

  
We summarize our main contributions.
  
\begin{enumerate}[leftmargin=*]
     \item We introduce the idea of "Code naturalization" as a pre-training task. 
     \item Using code from Github, and custom tooling, we have generated and  released a large dataset for pre-training models on the Naturalization task. 
     \item We have built and released a large Sequence-to-Sequence model pre-trained on Naturalization. 
     \item We show that (when appropriately fine-tuned) \tool outperforms SOTA on several settings.
     \end{enumerate}

We publish our source code and data download script for pre-training \tool anonymously in  \url{https://github.com/saikat107/NatGen.git}. 
We also share the pre-trained model in \url{https://bit.ly/3N0NGfG}. 
\section{Background \& Problem Formulation}
\label{sec:background}

This section presents the relevant technical background that leads to this work and an overview of the main research questions.

\subsection{\textbf{The Dual Channels of Code}}
Humans can read and write both natural languages and code. 
However, unlike natural language, source code involves 
\emph{two} channels of information: formal \& natural~\cite{casalnuovo2020does}. 
The formal channel, unique to code, affords precise, formal
semantics; interpreters, compilers, etc., use this channel.  
On the other hand, the natural channel (perhaps more probabilistic and noisy) relies on variable names, comments, etc., and is commonly used by humans for code comprehension and communication~\cite{casalnuovo2020programmers,casalnuovo2020does}. The formal channel's precision enables semantic preserving code transformation,  which supports static analysis, optimization, obfuscation, \emph{etc}.   For instance, major refactoring of a source code may drastically change the syntactic structure while preserving the semantics~\cite{fowler2018refactoring, ding2021contrastive}. However, not all the semantically equivalent code is ``natural"~\cite{hindle2016naturalness}---the usual way developers write code and thus, amenable to statistical models~\cite{hindle2016naturalness}. 
In fact, deviation from such ``naturalness" may lead to unintended bugs~\cite{ray2016naturalness}, and increase difficulty of human comprehension~\cite{casalnuovo2020programmers,casalnuovo2020does}. 

We leverage the natural/formal duality for our pre-training objective in this work. We keep the formal channel constant (not changing the meaning) for a given code and modify the syntax by creating ``unnatural'' code. Then we train the model to take the ``unnatural" code as input and do what a human Editor does with natural language text: understand the ``unnatural" code and generate more natural code that a developer would write. Thus, the model simultaneously learns to both 
comprehend code, \emph{and}  generate  ``natural'' code.

\subsection{\textbf{{``Naturalizing" \emph{vs.} De-noising}}} 
Naturalizing pre-training essentially follows in the tradition of 
\emph{denoising pre-training}, although, arguably, the former is more subtle and challenging. 
Denoising pre-training~\cite{lewis2019bart, lachaux2020unsupervised, ahmad2021unified} is a well-established pre-training strategy for encoder-decoder models: the encoder is presented with a noised-up input, and the decoder is asked to generate the original, noise-free input. By training the model to identify \& remove ``noise'' in a noisy output, 
(in theory) one teaches it to reason about and correctly generate text. Exactly what a model learns largely depends on the noise types. For instance, PLBART~\cite{ahmad2021unified} uses syntactic noise\footnote{Noise that breaks the syntax structure of code}(\ie token masking, token deletion, etc.). Thus, denoising pre-training enables  PLBART to learn both about the syntax of input source code, \emph{and} learn to generate syntactically correct code.
Naturalizing pre-training, on the other hand, begins with syntactically correct but artificially-created \emph{unnatural} source code and forces the model to generate correct \emph{semantically equivalent natural} code that is just what a human originally wrote. Such pre-training requires more subtle changes to the code. We hypothesize that this provides a more demanding pre-training setting, which will lead to better on-task code generation performance.

\subsection{Research Questions}
\label{subsec_rqs}

Our hypothesis is that our \emph{naturalizing} task (see \cref{sec:semantic_denoising}) 
endows our pre-trained model with the ability to generate syntactically {and semantically} correct, \emph{and} natural code. This
leads to several RQs. 

\RQ{1}{\rqa}

In contrast to existing de-noising techniques~\cite{ahmad2021unified} that help the model learn lexical \& syntactic structure, the naturalizing task, which is arguably more demanding than de-noising, forces \tool generating better code with higher syntactic and semantic correctness. 

The pre-training data we use (in \tool) challenges the model to naturalize code that was ``de-naturalized" in several ways, such as dead-code inserted, variable renamed, etc. We investigate the relative performance under different naturalization challenges. 

\RQ{2}{\rqb}

We evaluate the performance under different challenges on a held-out validation dataset. This dataset is sampled with the same distribution of de-naturalizing transforms as the training dataset ($\mathcal{D}_t$); on this set, the model to reconstruct the original code. 
Our exploratory investigation reveals that Variable Renaming is the hardest transformation to undo:  the model reconstructs original code with only $~$40\% accuracy. Dead Code, on the other hand, is the easiest with $~$99\% accuracy. 



We further investigate \tool's performance for downstream source code generation tasks.

\RQ{3}{\rqc}

We fine-tune the pre-trained \tool on task-specific training dataset for a certain time budget and evaluate the fine-tuned model on the benchmark testing dataset for corresponding task. These tasks include source code (java) generation from text, code translation (from Java to C\# and C\# to Java), and Bug fixing. After fine-tuning, \tool achieves the state-of-the-art performance in all these tasks. In addition, we also discover that, code generated by \tool are syntactically and semantically more closer to the expected code. 


We observe that training a model for a complex task requires sufficient labeled data. However, for most software engineering tasks, finding labeled data is a significant challenge~\cite{ahmad2021avatar}. We investigate potential scenario where size of the training data is extremely small. 

\RQ{4}{\rqd}

We simulate training data scarcity in two different ways -- {\em Zero-shot learning}, and {\em Few-shot learning}. 
For ``Zero-shot'' learning, we evaluate the pre-trained \tool in different tasks \emph{without} any task specific fine-tuning. 
For ``few-shot'' setting, we simulate training data scarcity by sub-sampling the benchmark training datasets. 
We fine-tune the pre-trained \tool on these limited training examples and measure the performance. 
We observe that 
\tool is very efficient in low-data training. Since \tool learns to generate syntactically and semantically correct code as part of pre-training, it faces less burden while learning in low-data training.



\section{Methodology}
\label{sec:method}

\lstset{escapechar=~,style=CustomJava}
\begin{figure*}[!htpb]
\centering
\begin{subfigure}{0.48\linewidth}
\begin{lstlisting}
int search(int[] arr, int key, int low, int high){
    while (low <= high) {
        int mid = low  + ((high - low) / 2);
        if(arr[mid] == key)  { return mid; }
        else { high = mid + 1; }
    }
    return -1;
}
\end{lstlisting}
\vspace{-5mm}
\caption{Original Code}
\label{fig:original_code}
\end{subfigure}%
\begin{subfigure}{0.48\linewidth}
\begin{lstlisting}
int search(int[] arr, int key, int low, int high){
    ~\red{\colorbox{highlight}{\bf for}}~(~\red{\colorbox{highlight}{\bf ;}}~low <= high~\red{\colorbox{highlight}{\bf ;}}~) {
        int mid = low  + ((high - low) / 2);
        if(arr[mid] == key) { return mid; }
        else { high = mid + 1; }
    }
    return -1;
}
\end{lstlisting}
\vspace{-5mm}
\caption{Loop Transformation}
\label{fig:loop_transformation}
\end{subfigure}
\vspace{2mm}

\rule{\textwidth}{0.1mm}
\begin{subfigure}{0.48\linewidth}
\begin{lstlisting}
int search(int[] arr, int key, int low, int high){
    while (low <= high) {
        int mid = low  + ((high - low) / 2);
        ~\red{\colorbox{highlight}{\bf while\quad(\quad i < \quad i \quad )\quad\{ }}~
            ~\red{\colorbox{highlight}{\bf high = mid + 1; }}~
        ~\red{\colorbox{highlight}{\bf\} }}~
        // ... Rest of the Code
    }
    return -1;
}
\end{lstlisting}
\vspace{-5mm}
\caption{DeadCode Insertion}
\label{fig:dead_code}
\end{subfigure}%
\begin{subfigure}{0.48\linewidth}
\begin{lstlisting}
int search(int[] arr, int key, int low, int high){
    while (~\red{\colorbox{highlight}{\bf high\quad>=\quad low}}~) {
        int mid = low  + ((high - low) / 2);
        if(~\red{\colorbox{highlight}{\bf arr$[$mid$]$\;\;$!=$\;\; key}}~)  { 
            ~\red{\colorbox{highlight}{\bf high\quad=\quad mid\;\;+\;\;1;}}~
        }
        else~\red{\colorbox{highlight}{
        \bf \{\;\;return\;\;mid;\;\;\}}}~
    }
    return -1;
}
\end{lstlisting}
\vspace{-5mm}
\caption{Block and Operand Swap}
\label{fig:block_swap}
\end{subfigure}
\vspace{2mm}

\rule{\textwidth}{0.1mm}

\begin{subfigure}{0.48\linewidth}
\begin{lstlisting}
int search(int[] arr, int key, int low, int high){
    while (low <= high) {
        int mid = low  + ((high - low) / 2);
        if(arr[mid] == key) { return mid; }
        else { 
            high = ~\red{\colorbox{highlight}{\bf mid+\;+}}~; 
        }
    }
    return -1;
}
\end{lstlisting}
\vspace{-5mm}
\caption{Inserting confusing code element}
\label{fig:confusion}
\end{subfigure}%
\begin{subfigure}{0.48\linewidth}
\begin{lstlisting}
int search(int[] ~\red{\colorbox{highlight}{\bf var\_1}}~, int key, int low, int ~\red{\colorbox{highlight}{\bf var\_2}}~){
    while (low <= ~\red{\colorbox{highlight}{\bf var\_2}}~) {
        int mid = low  + ((~\red{\colorbox{highlight}{\bf var\_2}}~ - low) / 2);
        if(~\red{\colorbox{highlight}{\bf var\_1}}~[mid] == key) { return mid; }
        else { ~\red{\colorbox{highlight}{\bf var\_2}}~ = mid + 1; }
    }
    return -1;
}
\end{lstlisting}
\vspace{-5mm}
\caption{Variable Renaming} 
\label{fig:var_renaming}
\end{subfigure}
\vspace{2mm}

\rule{\textwidth}{0.1mm}
\vspace{-6mm}
\caption{Semantic preserving transformation used to prepare the pre-training data for \tool.}
\label{fig:transformations}
\end{figure*}

Our approach comprises three steps: (i) ``De-Naturalize'' source code to accumulate pre-training data for \tool (\S\ref{sec:semantic_denoising}); (ii) pre-train \tool using this data for naturalization task (\S\ref{sec:pretrain_train}); (iii) Fine-tune pre-trained \tool with task specific dataset (\S\ref{sec:finetuning}). 


\subsection{De-Naturalizing Source Code}
\label{sec:semantic_denoising}

For the first step above, 
we use six rules to transform a natural code into its unnatural counterpart. These transformations are semantic-preserving but rewrite an original, natural, (human-) written code to an artificial form. Given a natural code element, we deploy an appropriate transformation, based on its AST structure and rewrite the code to ``de-naturalize'' it. 



\subsubsection{Designing Transformation Rules.}
\label{sec:data_prep}

We use six classes of de-naturalizing transformations. 
These transformations are motivated by prior work on functional reasoning about source code~\cite{ding2021contrastive, gopstein2018prevalence, gopstein2020thinking} and semantic bug-seeding~\cite{patra2021semantic}. 
~\Cref{fig:transformations} show the details. 

\textbf{Loop Transformation (\Cref{fig:loop_transformation}).}
This rule modifies {\tt for} loops into equivalent {\tt while} loop and vice-versa. We rewrite a {\tt while} loop of the form~\texttt{ while ( \highlight{condition} )}~\texttt{\{ loop-body \}} into a {\tt for} loop as ~\texttt{ for ( \highlight{\textbf{;}} \highlight{condition} \highlight{\textbf{;}} )}~\texttt{\{ loop-body \}}. 
Likewise, to transform a {\tt for} loop into a {\tt while} loop, we move the initializer of the {\tt for} (if any) before the loop, and the update {\em expression} (if any) of the {\tt for} loop as the last {\em statement} in the loop. We also add this update statement before any loop breaking statement (\ie \texttt{break}, \texttt{continue}). For example, we transform ``\texttt{for(\highlight{int i = 0;} i < 10; \highlight{i++})\{ if(i)\{ foo(); continue;\}  bar(); \}}'' as ``\texttt{\highlight{int i = 0;} while(i < 10)\{ if(i)\{ foo(); \highlight{i++;} continue;\}  bar(); \highlight{i++;}\}}''.

\textbf{Dead Code Injection (\Cref{fig:dead_code}).}
We inject blocks of dead code at random positions in the original code. 
By ``dead code" we mean code that appears in the source but is
never executed.
In \Cref{fig:dead_code}, we inject the code block \linecode{high = mid + 1;} at line 4 of the original code (\cref{fig:original_code}). To add challenge to the model, we transplant these inserted statements from the same input code. To ensure the "death" of inserted code, we put the inserted statements in a block headed by either a loop or a branch, guarded by a unsatisfiable condition
so that the code inside the block will never execute. In \Cref{fig:dead_code}, the condition \highlight{\tt i < i} is always \linecode{false}; and the code in line 5 is quite dead.

\textbf{Block Swap (\Cref{fig:block_swap}).}
Here we swap the ``then" block of a chosen {\tt if} statement with the corresponding {\tt else} block.  To preserve semantic equivalence, we negate the original branching condition. For instance, \cref{fig:block_swap} replaces the {\tt if} block (line 4 in \Cref{fig:original_code}) with the {\tt else} block (line 5 in \cref{fig:original_code}). We negate the original condition (\linecode{arr[mid] == key}) as (\linecode{arr[mid] != key}).

\textbf{Operand Swap (\Cref{fig:block_swap}).}
Here, we swap the operands of binary logical operations. For instance, we change the expression \linecode{low <= high} with \linecode{high >= low} in line 2 in \Cref{fig:block_swap}. When swapping the operands of a logical operator, we change the operator to make sure the modified expression is the logical equivalent to the one before modification. In case of asymmetric inequality operators ($>$, $<$, $>=$, $<=$), we change the direction -- keep as is for symmetric operators (\ie $==$, $!=$).

\textbf{Confusing Code Insertion (\Cref{fig:confusion}).}
We introduce confusing code patterns in the code as outlined by \citet{gopstein2018prevalence, gopstein2020thinking}. In particular, we introduce two forms of confusing code. First, we modify the of the form \linecode{\{i = j; j += 1;\}} to \linecode{i = j++;}. Second, we introduce ternary operator as applicable. For example, we transform the code \linecode{if (x != 0) \{y = p;\} else \{y = q;\}} to \linecode{y = (x != 0) ? p : q;}. 

\textbf{Variable Renaming (\Cref{fig:var_renaming}).}
We rename some variables to {\tt VAR\_i}. While renaming a variable, we analyze the dataflow of that variable and rename all occurrences of that variable in the entire code. From all the variables used in the code, we change just a certain percentage. For instance, in \Cref{fig:var_renaming}, we renamed variable \linecode{arr} to \linecode{var_1}, and variable \linecode{high} to \linecode{var_2}, leaving all other variables unchanged. Note that, unlike other transformations, variable renaming does not create AST of Dataflow graph difference. However, this challenging task~\cite{allamanis2017learning} forces the model to learn to generate natural variable names. 
This resembles the de-obfuscation pre-training task of~\cite{roziere2021dobf}. 

\begin{figure}[!ptb]
    \centering
    \includegraphics[width=\linewidth]{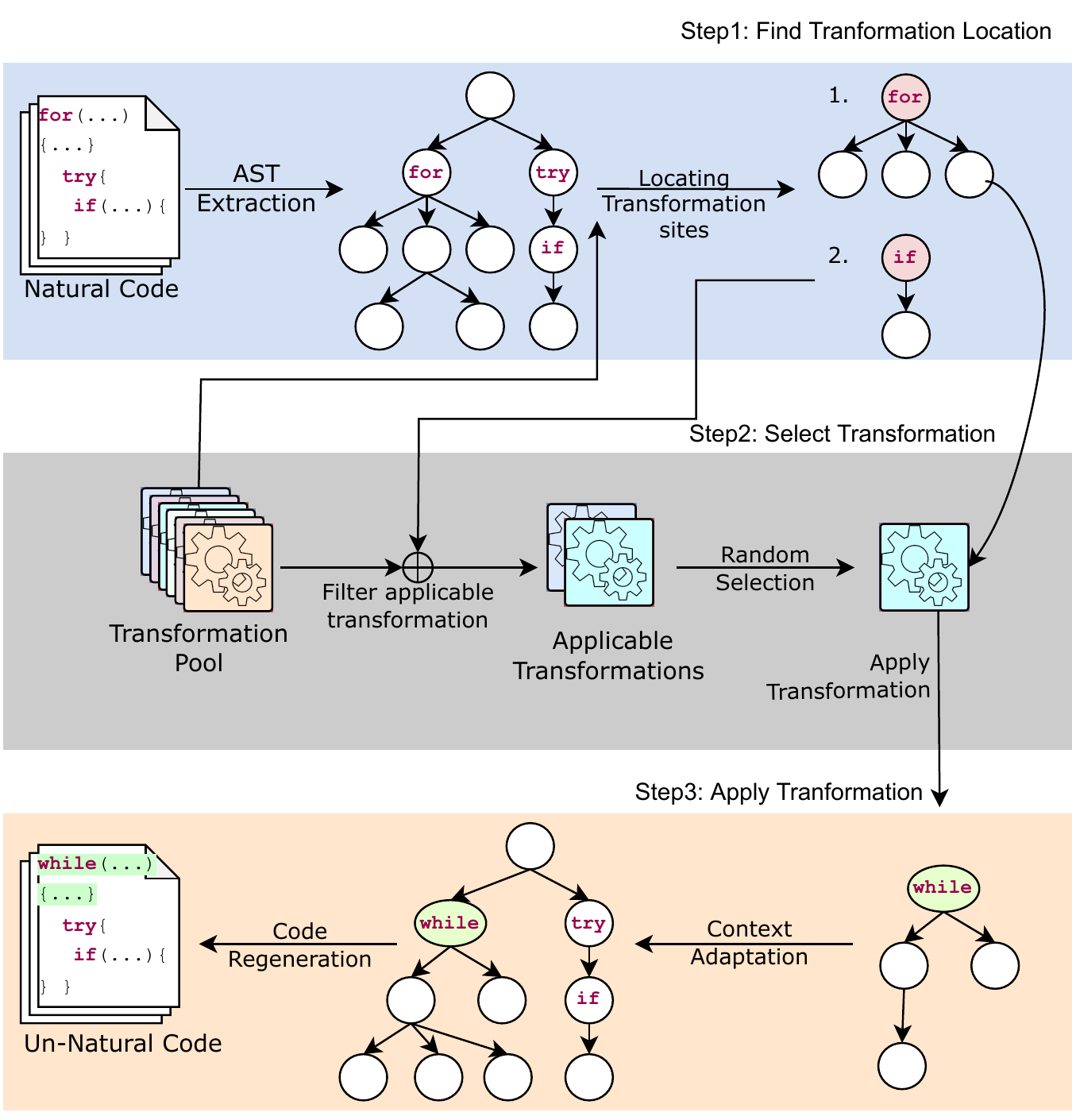}
    \vspace{-5mm}
    \caption{``De-Naturalization'' workflow in \tool.}
    \label{fig:approach}
\end{figure}

\subsubsection{Applying Transformation.}
\label{sec:transform}
Assume a set of transformation rules $\Phi = \{\phi_1, \phi_2,  \phi_3, ...\}$. 
Given  original code $c_i$, $\phi_j(c_i)$ transforms the code, changing the structure while preserving semantics. \Cref{fig:approach} shows how to apply such transformation to $c_i$. It works in three steps:
\begin{itemize}[leftmargin=*]
    \item {\em Find Transformation Location.} Given a piece of source code ($c_i$), we first use tree-sitter\footnote{\url{https://tree-sitter.github.io/tree-sitter/}}
    to parse out the AST ($T_{c_i}$). 
    From the AST, we extract potential locations for de-naturalization. These locations are nodes ($n_k$) in $T_{c_i}$. While choosing location $n_k$ from $T_{c_i}$, we consult $\Phi$ -- we extract the nodes where at least one of $\phi_j \in \Phi$ is applicable.
    
    \item {\em Select Transformation Rule.}  Once we have a set of such nodes, we filter out the transformation rules that cannot be applied to any node of in $T_{c_i}$. After such a filtration, we have a set of transformations $\Phi_a \subseteq \Phi$. At this stage, we randomly select one transformation pattern $\phi_j \in \Phi_a$ to apply at an application location (AST node) $n_k$.
    
    \item {\em Apply Transformation.} We apply $\phi_j$ to $n_k$ to get the transformed node $n'_{k}$. We then structurally match $n'_{k}$ with the original AST $T_{c_i}$, specifically $n_k$. We adapt the context of $n_k$ to the transformed node's ($n'_k$) context. In that way, we get the transformed AST ($T'_{c_i}$), which we then translate to get the transformed code $c'_i$. 
\end{itemize}

We designed the transformation function $\phi_j$ and subsequent context adaptation in such a way that preserves the meaning or functionality of the original code. We use AST analysis and (approximated) data flow analysis on code AST.

\subsection{Pre-training}
\label{sec:pretrain_train}
Once we have a pool of  ``unnatural'' code using the transformation in \Cref{sec:semantic_denoising} (\ie transform code $c_i$ as `un-natural' code $\phi_j(c_i)$),
we use a neural sequence-to-sequence translation model ($\mathcal{M}$) to reconstruct $c_i$ from $\phi(c_i)$, \ie we want $\mathcal{M}(\phi_j(c_i))$ to approximate $c_i$ . In particular, given a training dataset $\mathcal{D}_t = \{c_1, c_2, ...\}$ consisting of developers written code, set of   ``de-naturalizing'' transformations $\Phi = \{\phi_1, \phi_2, \phi_3, ...\}$, we optimize the following function to learn $\mathcal{M}$'s optimal parameter $\Theta$.
\begin{equation}
    \Theta = \argmin_{\theta} \sum_{c_i\in \mathcal{D}_t} CrossEntropy\left(\mathcal{M}\left(\phi_j\left(c_i\right)\right), c_i\right)
\end{equation}

\subsection{Fine-tuning}
\label{sec:finetuning}
The objective of our pre-training is to learn to both comprehend and generate general-purpose source code. However, different tasks related to source code generation (\eg text to code generation, code to code translation, bug fixing) call for task-specific training of the pre-trained model. This training phase on a pre-trained model is known as fine-tuning~\cite{devlin2018bert}. We consider the fine-tuning in \tool as a translation task and follow the standard transformer based-machine translation procedure~\cite{vaswani2017attention}. First, the encoder generates the encoded representation $R(X)$ given the input $X= [x_1, x_2, ..., x_n]$. The decoder then sequentially generates the output $Y = [y_1, y_2, ..., y_m]$. While encoding an input token $x_k$, the encoder learns the attention matrix \wrt every token in the input, including  $x_k$.
Such attention matrix is known as {\em self-attention}. While generating an output token $y_m$, the decoder learns  the attention matrix with all previously generated tokens $[y_1, y_2, ..., y_{m-1}]$ through {\em self-attention} and the encoder generated representation $R(X)$ through {\em cross-attention}. We refer to \citet{vaswani2017attention} for more detail about transformer-based translation.

\section{Experimental Setup}
\label{sec:experiemnts}

This section details the experimental design of \tool. 

\paragraph{\textbf{Pre-training data.}}
Following prior works~\cite{feng2020codebert, guo2020graphcodebert, wang2021codet5}, we primarily use CodeSearchNet~\cite{husain2019codesearchnet} dataset for the pre-training purpose. CodeSerachNet is a publicly available dataset with six languages: Java, Python, Go, JavaScript, Ruby, and PHP. In addition to CodeSearchNet, CodeT5 uses additional data for C and C\#.
We also use 1M functions each for C and C\#. For these two additional languages, we collected 5000 active projects from GitHub and randomly selected 1M functions considering the maximum sequence length of the model.
\begin{table}[!htb]
    \centering
    \small
    \caption{\small Statistics of fine-tuning datasets.}
    \vspace{-3mm}
     \resizebox{\linewidth}{!}{
        \begin{tabular}{lllrrr}
        \hlineB{2}
        & \textbf{Task} & \textbf{Dataset}  & \textbf{Train\#} & \textbf{Dev\#} & \textbf{Test\#} \bigstrut\\
        \hlineB{2}
        Text $\xrightarrow{}$ Code & Generation~\cite{iyer2018mapping} & Concode & 100000 & 2000 & 2000\bigstrut \\
        \hline
        Code $\xrightarrow{}$ Code & Translation~\cite{CodeXGLUE} & CodeXGLUE & {10300} & {500} & {1000} \bigstrut\\
        \hline
       \multirow{2}{*}{Text+code $\xrightarrow{}$ Code} & \multirow{2}{*}{BugFix~\cite{tufano2019learning}} & Small & 46628 & 5828 & 5831\bigstrut[t] \\
        & & Medium & 53324 & 6542 & 6538\bigstrut[b] \\
        \hlineB{2}
        \end{tabular}
     }
    \label{tab:finetune_data_stat}
\end{table}

\paragraph{\textbf{Fine-tuning data.}} 
We evaluate different variations of three benchmark tasks related to source code generation. The first task is {\em Text to Code generation}, where the input is an NL description of a Java method, and the output is the code. The second task is {\em Code Translation} between {\em Java to C\#} and {\em C\# to Java}. For this task, we evaluate Java-C\# parallel dataset proposed by ~\citet{CodeXGLUE}. The third and final task is {\em Bug Fix}, where the given a buggy code and a summary of the fix model generates the fixed code. For this task, we used the two different versions of the dataset (small, with less than 50 tokens and medium with up to 100 tokens) proposed by \citet{tufano2019learning}. Note that, similar to MODIT~\cite{chakraborty2021on}, we evaluate on {\em concrete} version of the refinement datasets.\Cref{tab:finetune_data_stat} shows the datasets and their statistics. For Text to Code Generation and Code Translation, we reuse the same split from CodeXGLUE~\cite{CodeXGLUE}, and for Bug Fix, we reuse the same split as MODIT. 

\paragraph{\textbf{Pre-training Model Configurations.}}
\label{sec:pretrain_exp}
We use 12 layer transformers with 12 attention heads on both encoder and decoder following the CodeT5~\cite{wang2021codet5} architecture. As discussed in \Cref{sec:method}, we use de-naturalization generative objectives for pre-training. 
We initialize our model with CodeT5's~\cite{wang2021codet5} released parameters. 
In particular, we initialize \tool with ``CodeT5-base'' model. 
We pre-train \tool on 2 Nvidia GeForce RTX 3090 GPUs for 25K steps, maintaining the effective batch size at 1080 with learning rate 5e-5. We train \tool for approximately 168 hours.

\paragraph{\textbf{Evaluation Metric.}}
\label{sec:metric}
Throughout the experiments in this work, we  evaluate accuracies w.r.t. exact match (EM), Syntax match (SM), Dataflow match (DM), and CodeBLEU (CB)~\cite{ren2020codebleu}. 
SM is the proportion of matching subtrees between output code and tadget code's ASTs \wrt number of all possible subtrees in the target code's AST. 
DM is the percentage of matched (with target code) anonymized dataflow edge (def-use edge) of output code \wrt all dataflow edges in the target code. 
Note that, both the SM and DM are components of CB. We explicitly evaluate these for understanding the syntactic and semantic correctness of generated code. We reuse Microsoft CodeXGLUE tool~\cite{codebleu_calculator} to compute SM, DM, and CB.  

\paragraph{\textbf{Baselines.}}
\label{sec:common_baselines}
While comparing the evaluation results for different tasks, we compare with large scale pre-trained models, including GPT-2~\cite{radford2019language}, CodeGPT~\cite{CodeXGLUE}, PLBART~\cite{ahmad2021unified}, SPT-Code~\cite{niu2022spt} and CodeT5~\cite{wang2021codet5}. Most of our fine-tuning evaluation is on benchmarked dataset; thus, we report the available results from CodeXGLUE leaderboard~\cite{cxg_leaderboard}. There are some task specific baselines, which we discuss while describing corresponding task.

\section{Empirical Results}
\label{sec:result}

We evaluate \tool on (i) pre-training and (ii) three fine-tuning tasks. We also check \tool's effectiveness in zero-shot and few-shot settings.

\subsection{\tool's Effectiveness on pre-training}
\label{sec:rq1}

\RQrepeat{1}{\rqa}

\paragraph{\underline{Motivation.}} 
We investigate whether pre-training on naturalizing task helps the model generate  correct and natural code
(code that is syntactically and semantically similar to the original code). 

\paragraph{\underline{Experimental Setup.}} 
We compare three large scale pre-trained models: (i) CodeT5~\cite{wang2021codet5}, (ii) PLBART~\cite{ahmad2021unified}, and (iii) \tool.
Note that, since PLBART is only pre-trained on Java and Python, we compare PLBART only for those languages, with the corresponding results of other models. 
We ask each of these models to reconstruct developers' written code from its de-naturalized (but semantically identical, see \S\ref{sec:semantic_denoising} \& \S\ref{sec:data_prep}) variants. We use the held-out validation data from our training procedure for this evaluation.
We evaluate the models for generating the Exact Match (EM), Syntax Match (SM) and Dataflow Match (DM). 

\begin{table}[!htb]
{
    \small
    \centering
    \caption{Evaluation of \tool for code generation task. CS is the percentage of examples where output is directly copied from source, and ED is the median edit distance between input code and output code.}
    \vspace{-3mm}
    \resizebox{\linewidth}{!}
    {
        \begin{tabular}{l|l|rrrr|rr}
            \hlineB{2}
            \textbf{Eval Data} & \textbf{Model} &  \textbf{EM} & \textbf{SM} & \textbf{DM} & \textbf{CB} & \textbf{CS} & \textbf{ED} \bigstrut\\
            \hlineB{2}
            \multirow{2}{*}{Full} & CodeT5 & 0 & 13.93 & 19.86 & 9.74 & 0\% & {60}  \bigstrut[t]\\
            & \tool & \textbf{70.39} & \textbf{98.78} & \textbf{97.69} & \textbf{97.31} & 0.01\% & 8 \bigstrut[t]\\
            \hlineB{2}
            \multirow{3}{*}{Java \& Py} & CodeT5 & 0 & 13.83 & 23.67 & 10.87 & 0\% & {65} \bigstrut[t] \\
            & PLBART & 0 & 73.17 & 75.95 & 74.56 & {7.05\%} & 3 \bigstrut\\
            & \tool & 64.13 & 98.16 & 96.85 & 96.82 & 0.01\% & 10\bigstrut[b]\\
            \hlineB{2}
        \end{tabular}
    }
    \\
    \label{tab:pretrain_eval_full}
}
\end{table}

\paragraph{\underline{Results.}} 
\Cref{tab:pretrain_eval_full} shows the evaluation results. 


\noindent
$\boldsymbol{\cdot}$~\textit{Syntax Match.} We find that the code generated by PLBART and \tool are mostly syntactically correct. However, CodeT5's does not always generate syntactically valid code, suggesting an advantage for naturalization pre-training. 
For instance, \Cref{fig:example_code_for_pretraining} shows code generated by different models from the given input. 
As we can see, CodeT5 generates a syntactically erroneous fragment. In contrast, PLBART made a minor edit on the input code, just removing the \linecode{protected} keyword. Both PLBART and \tool are pre-trained to generate complete code rather than fragments (which is the case of CodeT5~\cite{raffel2019exploring}); thus, the former two generally do better at generating syntactically correct code.

\lstset{escapechar=@,style=CustomJavaWoNumbers}
\begin{figure}[!htb]
    \centering
{
\begin{tabular}{p{0.46\linewidth}|p{0.46\linewidth}}
\begin{center}
    \footnotesize \underline{1. Input}
\end{center}
\begin{lstlisting}
protected SDV iam(SDV in,...){ 
    if(i < i){ 
        return new IAM(...); 
    } 
    return new IAM(...); 
}
\end{lstlisting}    &  
\begin{center}
    \footnotesize  \underline{2. PLBART output}
\end{center}
\begin{lstlisting}
SDV iam(SDV in, ...){ 
    if(i < i){ 
        return new IAM(...); 
    } 
    return new IAM(...); 
}
\end{lstlisting}\vspace{1mm}
\\
\hlineB{1}
\begin{center}
    \footnotesize  \underline{3. \tool output}
\end{center}
\begin{lstlisting}
protected SDV iam(SDV in,...){ 
    return new IAM(...); 
}
\end{lstlisting}    &  
\begin{center}
    \footnotesize  \underline{4. CodeT5 output}
\end{center}
\begin{lstlisting}
if (in) {
    @\errorcode{{\textbf{return} ~~~}}@
} @\errorcode{\}}@
\end{lstlisting}\\
\end{tabular}
}
\vspace{-4mm}
 \caption{Example of input generated code by different pretrained models (slightly simplified).}
\label{fig:example_code_for_pretraining}
\end{figure}


\noindent
$\boldsymbol{\cdot}$~\textit{Semantic Match.}
\tool is effective at recovering developers' written 
code from its de-naturalized semantic variants---around 70\% of the generated code (CodeBlue = 97\%) {\em exactly matches} the original code. 
PLBART, which deploys syntactic denoising, is at the second position in terms of CodeBlue.  

\tool also dominates the other two models in generating syntactically (SM) \& semantically (DM) valid code. 
While PLBART appears to generate syntactically correct code, it mostly copies code from the input---median edit distance from PLBART's input and the generated code is 3 (see \Cref{tab:pretrain_eval_full}). In fact, in 7.05\% of cases, PLBART just copies the input! By contrast, \tool learns to generate \emph{variants} of the input code, with only 0.01\% direct copy and a median edit distance of 10. Since PLBART is trained to remove syntax errors from the input, we conjecture that it does not inherently learn the semantic variation of the code. By contrast, we expose \tool to semantic code variations, forcing it to learn to generate code that is both more natural \underline{\emph{and}} semantically equivalent. 



$\boldsymbol{\cdot}$~\textit{Closer look into CodeT5.} 
Unlike \tool and PLBART, CodeT5 is not explicitly trained to generate complete code. During pre-training, CodeT5 learned to ``unmask'' masked token sequences. Thus, to better measure CodeT5's generation capacity, we conduct another experiment where we replaced all occurrences of some of the variable names in code with a special {\tt MASK1}, {\tt MASK2} tokens and asked CodeT5 to generate. This is one of the objectives (masked identifiers prediction) CodeT5 is pre-trained to optimize. We take the CodeT5's output and identify all potential identifiers~\footnote{we use regex \linecode{"[A-Za-z\_]+[A-Za-z0-9_]*"} to find identifiers.}. Surprisingly, in {\em only} 0.27\% of the cases, could CodeT5  generate {\em all} the variables, and in 0.61\% of cases {\em half} of the masked variables., while \tool successfully translates 40.45\% of those examples back to its original code, including correctly predicting the replaced variable names. In addition, CodeT5's generated token sequence contained a lot of other tokens than the variable names (\cref{fig:example_code_for_pretraining}.4, for example).


\RS{1}{
    Naturalization enables \tool to reason about code semantics and thus help generate more natural code variants than existing pre-training models and pre-training objectives.
}


We also did an ablation study evaluating the effect of \tool's different components on the results.  

\RQrepeat{2}{\rqb}
\paragraph{\underline{Motivation.}}  
\label{sec:rq2}


In this RQ, we study how different transformation rules (see \S\ref{sec:semantic_denoising})
contribute to learn generating natural code from  different semantic variants . 
We also evaluate how well \tool learns that in different programming languages over training time.

\paragraph{\underline{Experimental Setup.}}  While pre-training, we checkpoint the \tool model every 1k training steps,  for a full run of 25k steps. At each checkpoint, we evaluate the naturalization task performance. Before training, we held out 0.1\% of the total data as validation data. Note that, since our goal in this experiment is to understand \tool's pre-training better, we ``de-naturalized" the validation data using the same training data distribution. This setting gives us a controlled environment for experimentation.

\begin{figure}[!tbh]
    \centering
    \includegraphics[width=\linewidth]{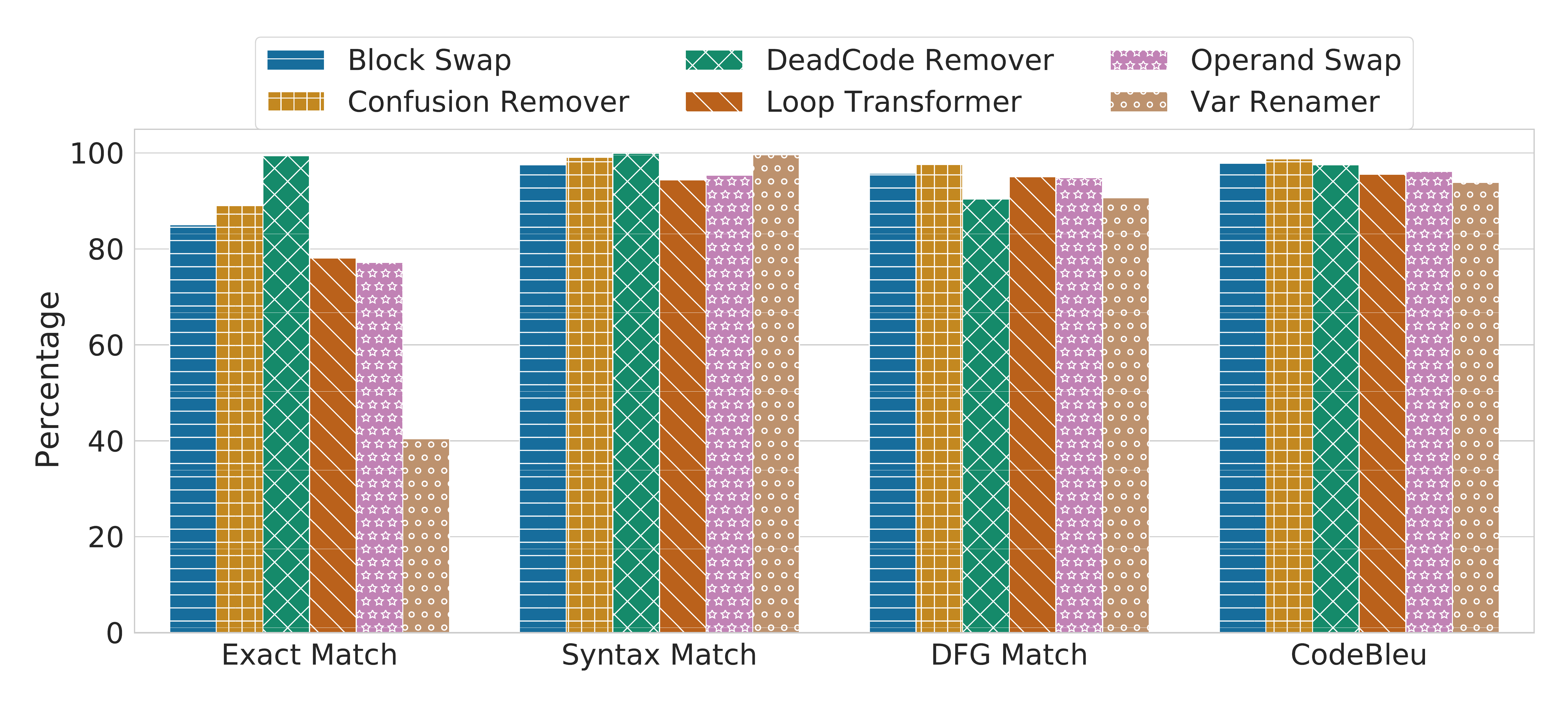}
    \vspace{-8mm}
    \caption{Performance of \tool pre-trained model under different code transformations.}
    \label{fig:transfromation_performance}
\end{figure}

\paragraph{\underline{Results.}} 

\Cref{fig:transfromation_performance} shows \tool's performance under different types of semantic variants. 
Results show that \tool  has most trouble recreating the original code (just 40\% Exact Match) with the variable renaming task. 
Variable renaming is challenging even for human developers~\cite{allamanis2015suggesting}---different developers may propose different names for the same object. Nevertheless, on this task, \tool achieves good syntax and dataflow match (99\% and 92\% respectively), indicating that \tool preserves syntax or semantics in most cases while generating code with renamed variables.

On the other hand, \tool can eliminate Dead Code 
with 99\% accuracy.  This result may be an artifact of our specific implementation of this transformation. Our dead-code insertion rule is simple, and formulaic; so the \tool quickly learns to identify and remove such dead code. A more complex pattern of dead code may challenge the model more, and help make it more robust; we leave this for future work. For naturalizing other transformations, \tool achieves more than 80\% exact match accuracy for Block swap and Confusion removing, and more than 75\% exact match accuracy for the rest. In all cases, syntax match, dataflow match, and CodeBLEU are well above 90\%. 



\begin{figure}[!htb]
    \centering
    \includegraphics[width=.65\linewidth]{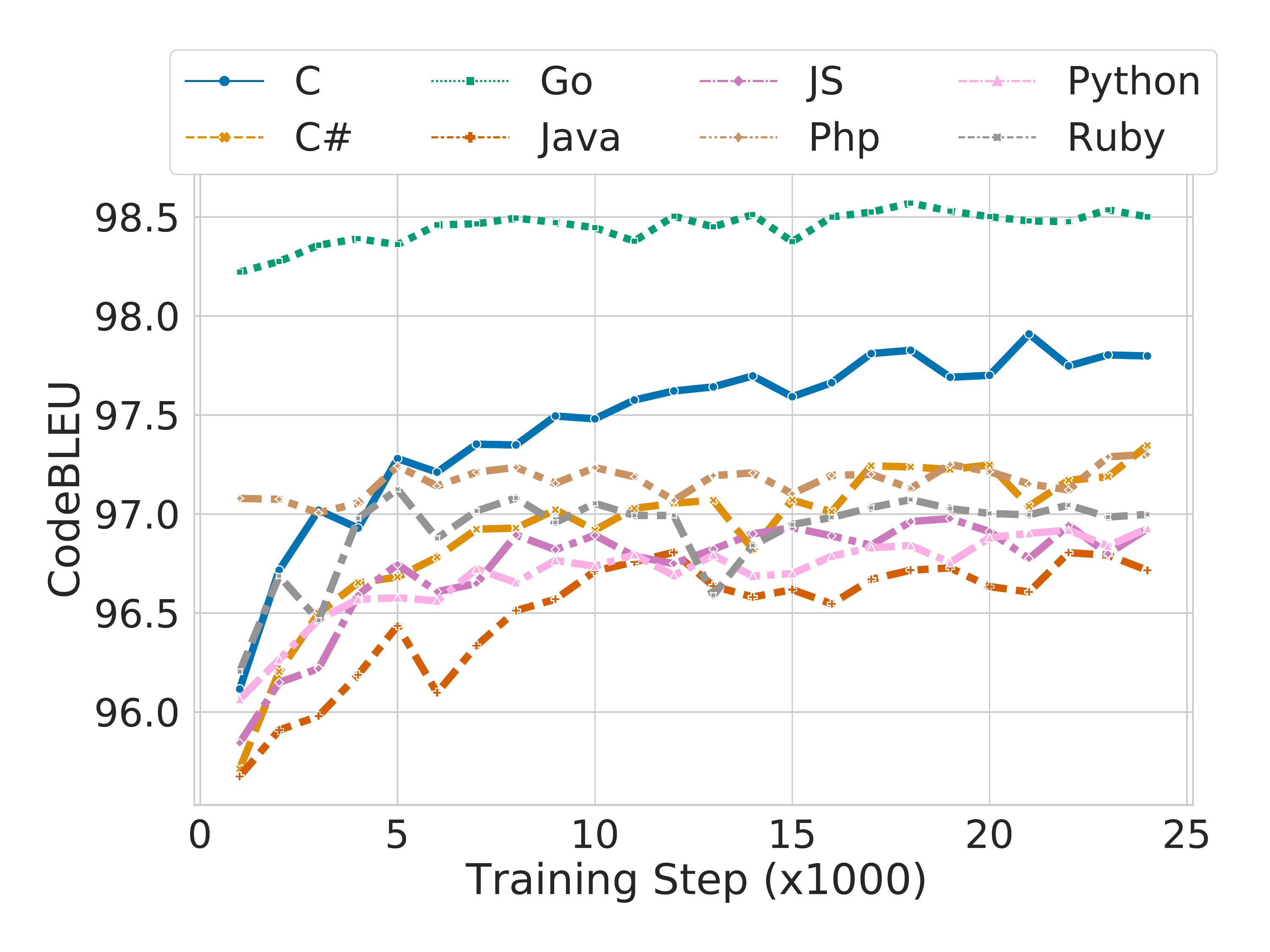}
    \vspace{-5mm}
    \caption{Progression of CodeBLEU of different language in Validation dataset over number pre-training steps.}
    \label{fig:pretrain_perf_over_step}
\end{figure}

 \Cref{fig:pretrain_perf_over_step} shows how validation performance improves for different languages, with more training steps. Across all the languages the performance rapidly increases over the first few thousand training steps. In fact, at the beginning of (step 0) of \tool's pre-training, the overall exact match is 0, syntax match is 13.93\%, dataflow match is 19.86\%  and CodeBLEU is 9.74\% (see \Cref{tab:pretrain_eval_full} for details\footnote{
 \tool's pre-training start from CodeT5-base. Thus, 
 CodeT5-base is \tool's checkpoint at step 0.}). However, after just 1000 steps of training, the exact match rises to 61\%, syntax match to 97\%, dataflow match to 94\%, and CodeBLEU to 95\%. These metrics continue improving as  training progresses. These results confirm that across all the languages \tool gradually learns to generate more natural code from semantic variants.

\RS{2}{
pre-training performance depends on the types of semantic variants---while variable renaming seems the most difficult ($\sim$40\% accuracy), dead-code elimination appears to be an easier task ($\sim$99\% accuracy) to learn.
}
\subsection{\tool's Effectiveness on Fine-tuning Tasks}
\label{sec:rq3}

This section evaluates \tool's performance on three benchmark source code generative tasks. 

\RQrepeat{3}{\rqc}

\begin{table}[!htb]
    \small
    \caption{Results of Text to Code Generation. `-' implies that those results are not reported by corresponding approaches. $\mathcal{M}_{last}$ is the model after completing the fintuning, and $\mathcal{M}_{best}$ is the intermediate model with best validation performance.
    }
    \vspace{-0.2cm}
    \centering
    \begin{tabular}{rl|cccc}
        \hlineB{2}
        \multicolumn{2}{c|}{\textbf{Approach}} & \textbf{EM} & \textbf{SM} & \textbf{DM} & \textbf{CB} \bigstrut\\
        \hlineB{2}
        \multicolumn{2}{c|}{Seq2Seq} &  3.05 & - & - & 26.39  \bigstrut[t]\\
        \multicolumn{2}{c|}{\citet{guo-etal-2019-coupling}} &  10.05 & - & - & 29.46  \\
        \multicolumn{2}{c|}{\citet{iyer2018mapping}} &  12.20 & - & - & -\\ 
        \multicolumn{2}{c|}{GPT-2} &  17.30 & - & - & 29.69 \\
        \multicolumn{2}{c|}{CodeGPT} &  20.10 & - & - & 35.98 \\
        \multicolumn{2}{c|}{PLBART} &  18.75 & - & - & 38.52 \\
        \hline
        \multicolumn{2}{c|}{CodeT5-base} & \multirow{2}{*}{\textbf{22.30}} &  \multirow{2}{*}{-} & \multirow{2}{*}{-} & \multirow{2}{*}{43.20} \bigstrut[t]\\
        \multicolumn{2}{c|}{(reported)} & \bigstrut[b]\\
        \hline
        \multirow{2}{*}{CodeT5*} & $\mathcal{M}_{last}$ & 21.85 & 44.34 & 44.52 & 41.75 \bigstrut[t]\\
        & $\mathcal{M}_{best}$ & 21.55 & 41.08 & 43.71 & 38.30 \bigstrut[b]\\
        \hline
        \multirow{2}{*}{\tool} & $\mathcal{M}_{last}$ &  22.25 & \textbf{45.59} & \textbf{46.87} & \textbf{43.73} \bigstrut[t] \\
         & $\mathcal{M}_{best}$  & \textbf{22.30} & 44.38 & 45.64 & 42.44 \bigstrut[b]\\
         \hlineB{2}
    \end{tabular}
    
    {\footnotesize {* Our reproduced result using CodeT5's publicly available pre-trained model.}}
    \label{tab:code_generation}
\end{table}

\paragraph{\underline{Baselines.}} 
In addition to the baselines discussed in \Cref{sec:common_baselines}, for the {\em Text to Java Code generation} task, we compare with a group of baselines with no pre-training involved. 
These baselines include LSTM based Sequence to sequence models, \citet{guo-etal-2019-coupling}'s, and \citet{iyer2018mapping}'s proposed techniques. 
We also report our reproduced version of CodeT5 results in different tasks, slightly different from what they reported. 
For both the {\em Bug Fix} task, we compare with the reported results of MODIT~\cite{chakraborty2021on} and our reproduced CodeT5 result. 
\begin{figure*}[!tbh]
    \centering
    \begin{subfigure}{0.25\textwidth}
    \centering
    \includegraphics[width=\linewidth]{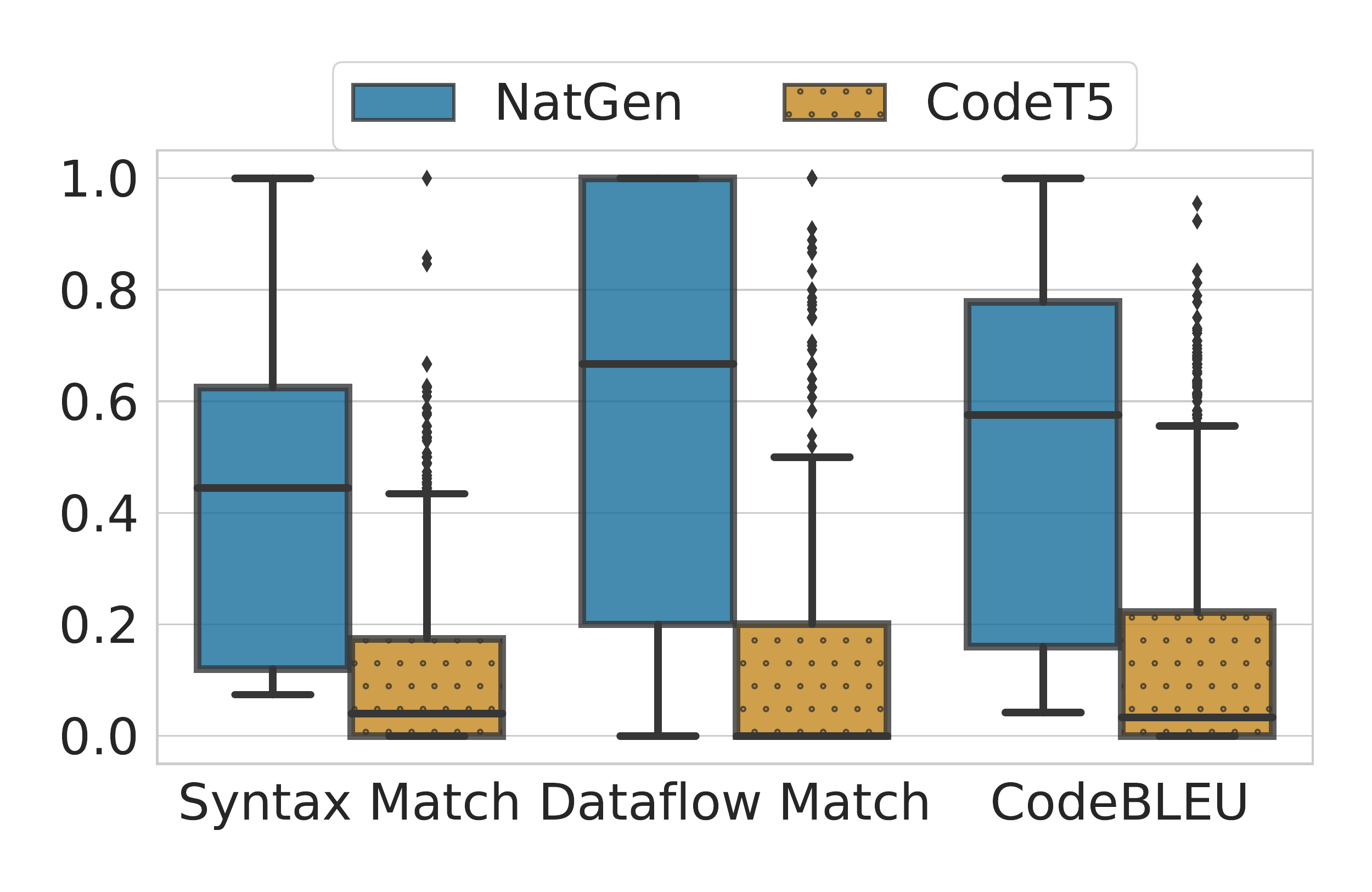}
    \vspace{-8mm}
    \caption{\footnotesize Java to C\# Translation}
    \label{fig:java-cs-zero-shot}
    \end{subfigure}%
    \begin{subfigure}{0.25\textwidth}
    \centering
    \includegraphics[width=\linewidth]{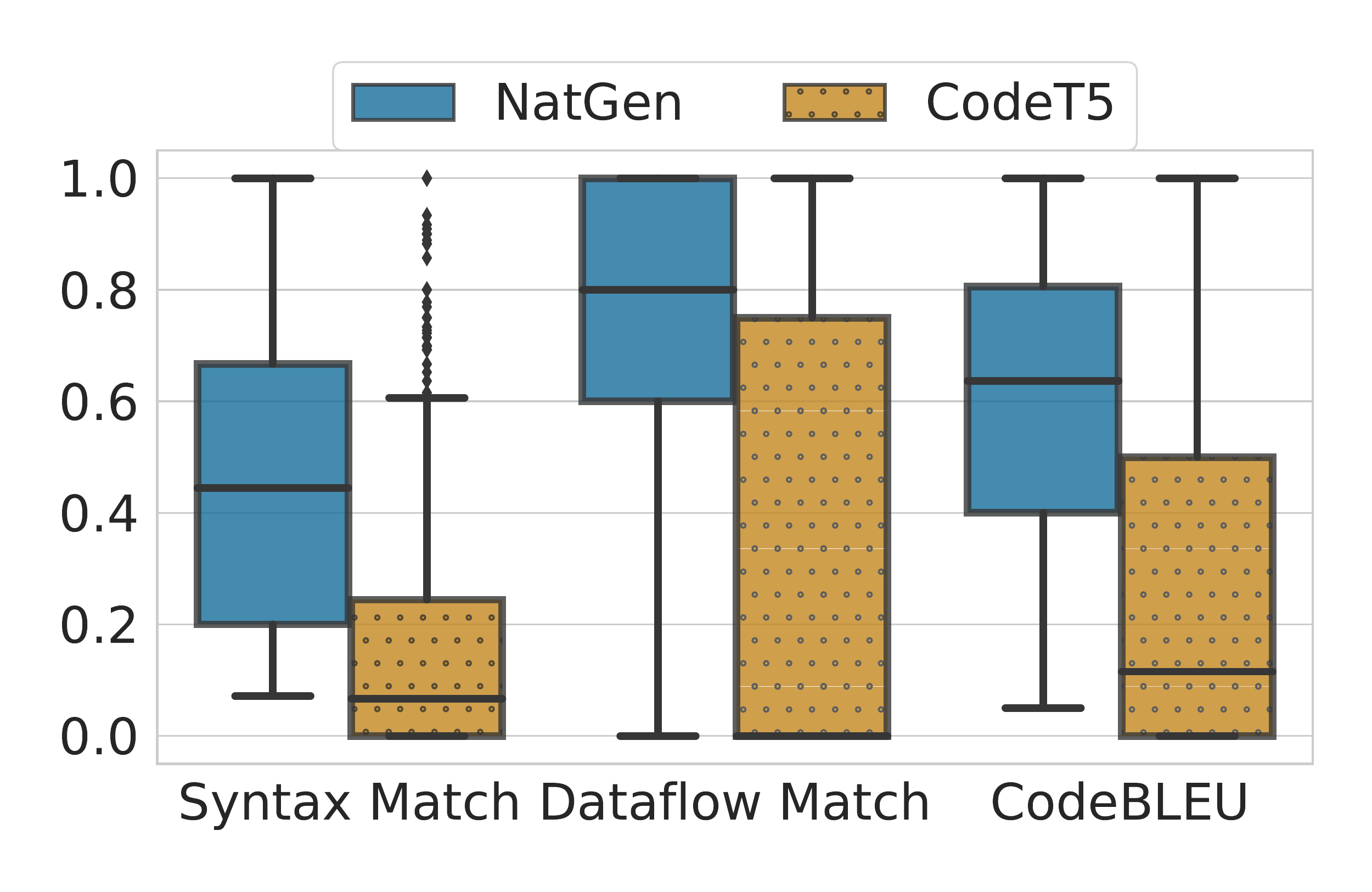}
    \vspace{-8mm}
    \caption{C\# to Java Translation}
    \label{fig:cs-java-zero-shot}
    \end{subfigure}%
    \begin{subfigure}{0.25\textwidth}
    \centering
    \includegraphics[width=\linewidth]{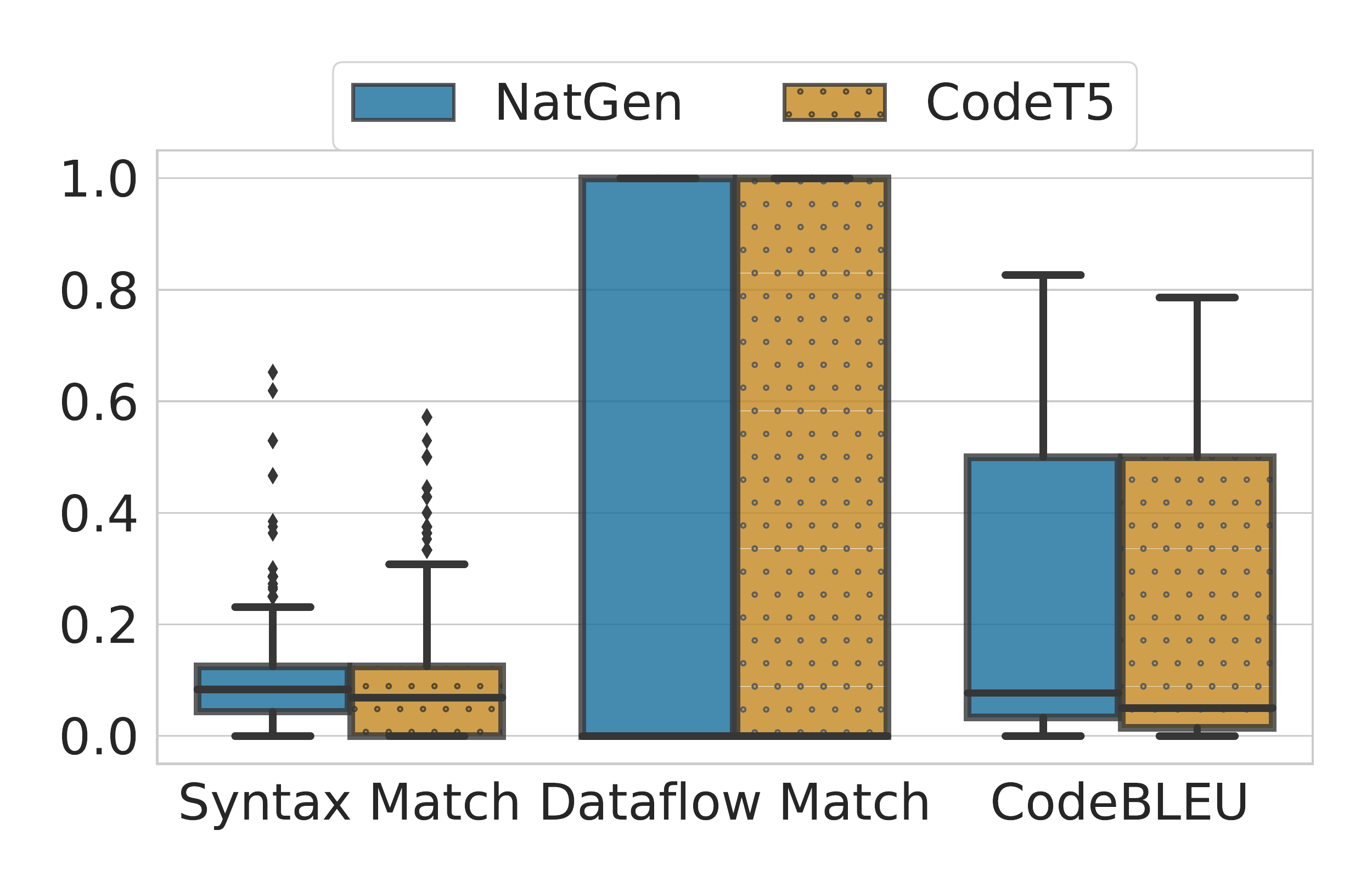}
    \vspace{-8mm}
    \caption{\footnotesize Text to Code Generation}
    \label{fig:concode-zero-shot}
    \end{subfigure}%
    \begin{subfigure}{0.25\textwidth}
    \centering
    \includegraphics[width=\linewidth]{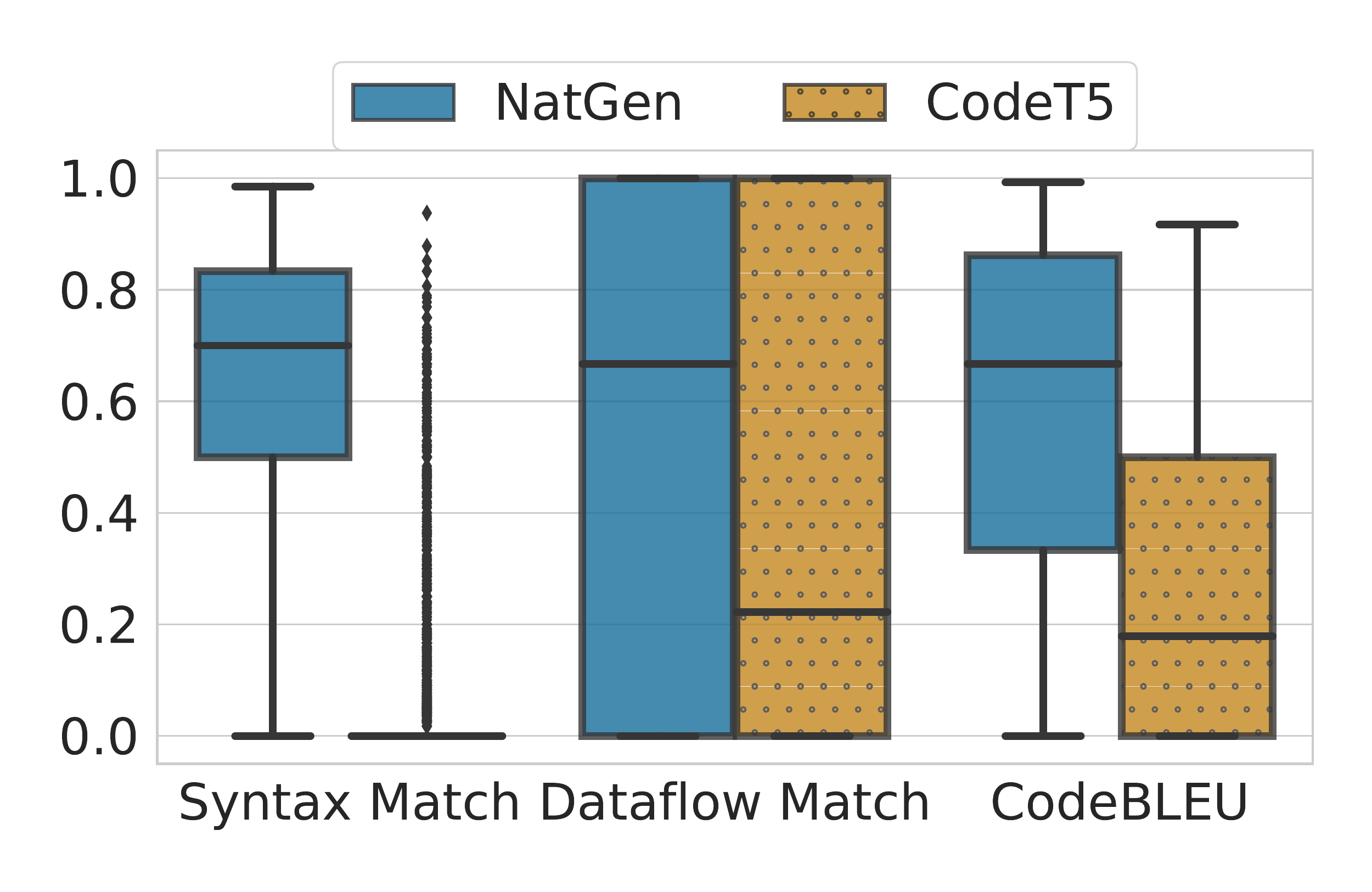}
    \vspace{-8mm}
    \caption{\footnotesize  Bug Fix (small, multimodal)}
    \label{fig:refine-zero-shot}
    \end{subfigure}
    \vspace{-3mm}
    \caption{Zero-shot transfer learning capability of \tool in for different tasks.}
    \label{fig:zero-shot-boxplot}
\end{figure*}

\paragraph{\underline{Results.}}~~\\

\vspace{-3mm}
\noindent\textbf{Text to Code Generation.}  
\Cref{tab:code_generation} shows evaluation results for text to code generation. 
We trained for 30 epochs. We stopped the training is the validation performance does not increase for more than three(3) consecutive epochs. For both CodeT5 and \tool, we report the performance of final model after the fine-tuning terminated ($\mathcal{M}_{last}$) and the performance of the model with best validation perfomance ($\mathcal{M}_{best}$). 
Interestingly, for both CodeT5 and \tool, the $\mathcal{M}_{last}$ model performs better than the corresponding $\mathcal{M}_{best}$ model. 
The result shows that \tool's generated code are more syntactically and semantically closer to the target code. 
The $\mathcal{M}_{last}$ model of \tool outperforms CodeT5's $\mathcal{M}_{last}$ model by 2.8\% in SM, 5.28\% in DM and 4.74\% in CB.
We conjecture that \tool's pre-training with ``naturalization'' help generate more natural code. 

\begin{table}[!htb]
    \caption{Code Translation results. `-' implies that those results are not reported by corresponding approaches.}
    \vspace{-0.2cm}
    \centering
    \resizebox{\linewidth}{!}
    {
    \begin{tabular}{c|cccc|cccc}
    \hlineB{2}
    \multirow{2}{*}{\textbf{Approach}} & \multicolumn{4}{c|}{\textbf{Java~ $\xrightarrow{}$~C\#}} & \multicolumn{4}{c}{\textbf{C\#~ $\xrightarrow{}$~Java}}\bigstrut\\
    \cline{2-9}
    & \textbf{EM} & \textbf{SM} & \textbf{DM} & \textbf{CB} &  \textbf{EM} & \textbf{SM} & \textbf{DM} & \textbf{CB} \bigstrut\\
    \hlineB{2}
    PBSTM & 12.5 & - & - & 42.7 & 16.1 & - & - & 43.5 \bigstrut[t] \\
    CodeBERT & 59.0 & - & - & 85.1 & 58.8 & - & - & 79.4 \\
    SPT-Code &  64.1 & - &- & - & 60.2 & - & - & - \\
    PLBART & 64.6 & - & - & 87.9 & 65.0 & - & - & \textbf{85.3} \bigstrut[b]\\
    \hline
    CodeT5 & \multirow{2}{*}{65.9} & \multirow{2}{*}{-} & \multirow{2}{*}{-} & \multirow{2}{*}{-} & \multirow{2}{*}{66.9} & \multirow{2}{*}{-} & \multirow{2}{*}{-} & \multirow{2}{*}{-} \\
    (reported) & & & & \\
    \hline
    CodeT5* & 65.9 & 90.4 & 91.9 & 87.8 & 66.0 & 90.4 & 88.9 & 84.4 \bigstrut \\
    \hline
    \tool & \textbf{66.2} & \textbf{91.0} & \textbf{92.0} & \textbf{88.1} & \textbf{67.3} & \textbf{91.0} & \textbf{89.8} & 85.2 \bigstrut \\
    \hlineB{2}
    \end{tabular}
    }
    {\footnotesize {* Our reproduced result using CodeT5's publicly available pre-trained model.}}
    \label{tab:code_translation}
\end{table}

\noindent\textbf{Code Translation.} 
\Cref{tab:code_translation} shows the results of \tool and different baselines for Code Translation. 
For Java to C\# translation, \tool achieves exact match accuracy of 66.2\% while CodeT5's accuracy is 65.9\%. 
In C\# to Java translation, \tool achieves 67.3\% exact match accuracy, which CodeT5 achieves 66.0\%. 
In addition, the syntactic match (SM), Dataflow match, and CodeBLEU are also higher than that of CodeT5. 

\begin{table}[!h]
    \caption{Result of Bug fix (Top 1 fix accuracy).}
    \label{tab:code_refinement}
    \vspace{-3mm}
    \centering
    \small
    \begin{tabular}{c|c|c|c|c}
    \hlineB{2}
    \multirow{2}{*}{\textbf{Approach}} & \multicolumn{2}{c|}{\textbf{BugFix$_{small}$}} & \multicolumn{2}{c}{\textbf{BugFix$_{medium}$}} \bigstrut \\
    \cline{2-5}
    & \textbf{Unimodal} & \textbf{Multimodal} & \textbf{Unimodal} & \textbf{Multimodal} \bigstrut\\
    \hlineB{2}
    MODIT & 20.35 & 21.57 & 8.35 & 13.18 \bigstrut[t] \bigstrut\\
    CodeT5 & 21.79 & 22.97 & 12.59 & \textbf{14.94} \bigstrut[b]\\
    \hline
    \multirow{1}{*}{\tool} & \textbf{22.26} & \textbf{23.43} & \textbf{13.32} & 14.93 \bigstrut[t] \bigstrut\\
    \hlineB{2}
    \end{tabular}
\end{table}

\begin{figure*}[!tbh]
    \centering
    \begin{subfigure}{0.25\textwidth}
    \centering
    \includegraphics[width=\linewidth]{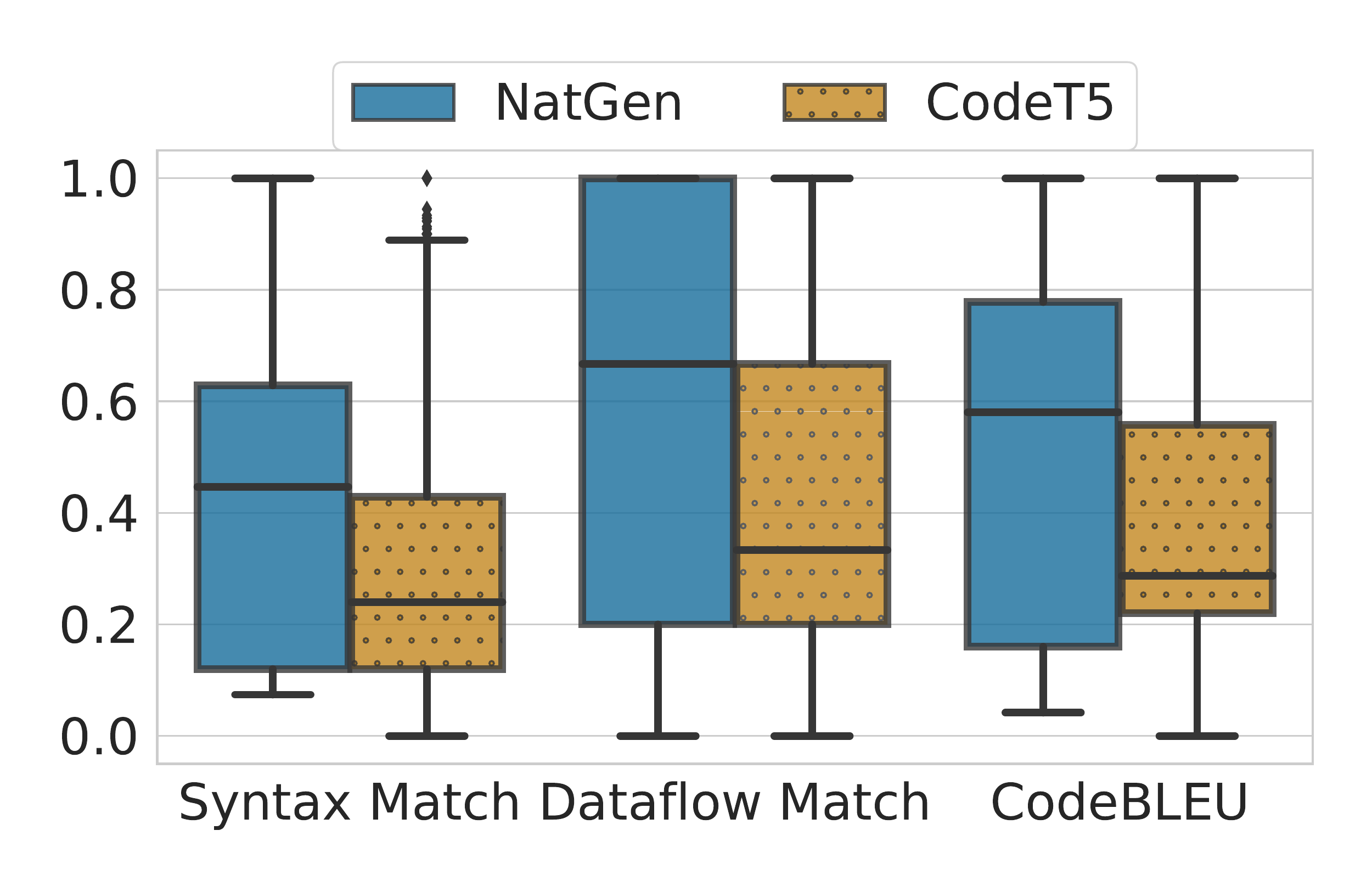}
    \vspace{-8mm}
    \caption{\footnotesize Java to C\# Translation}
    \label{fig:java-cs-few-shot}
    \end{subfigure}%
    \begin{subfigure}{0.25\textwidth}
    \centering
    \includegraphics[width=\linewidth]{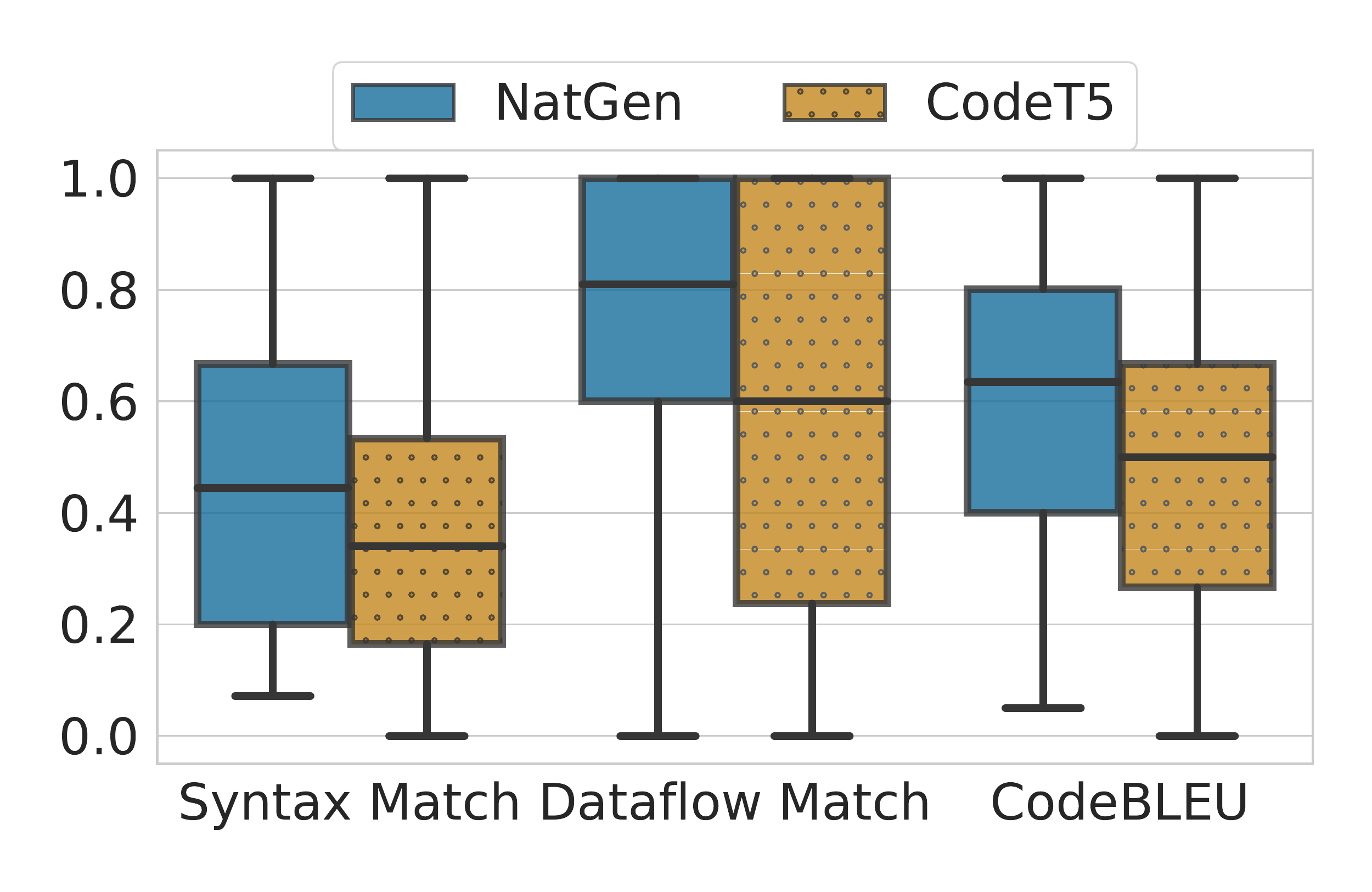}
    \vspace{-8mm}
    \caption{\footnotesize C\# to Java Translation}
    \label{fig:cs-java-few-shot}
    \end{subfigure}%
    \begin{subfigure}{0.25\textwidth}
    \centering
    \includegraphics[width=\linewidth]{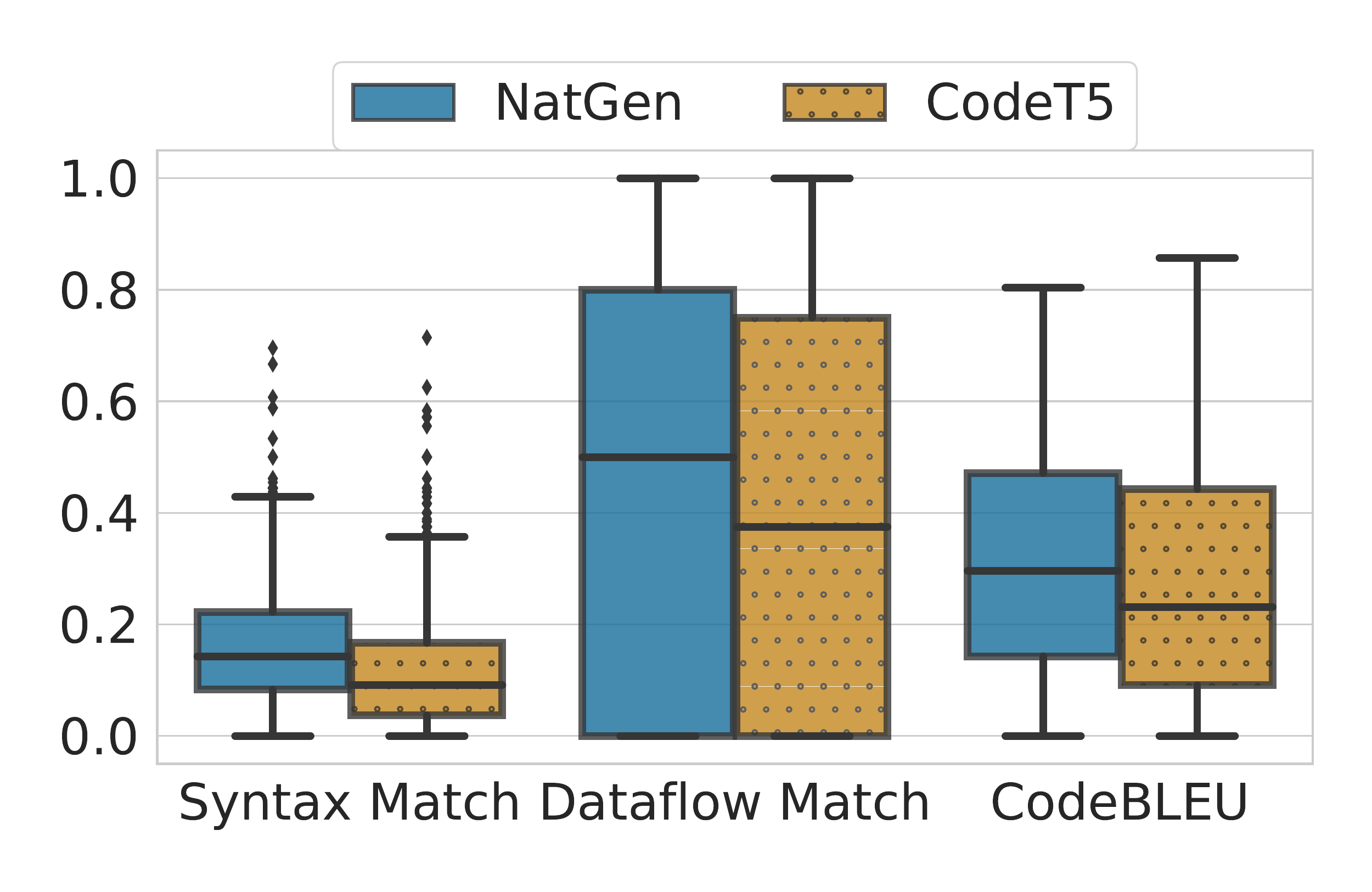}
    \vspace{-8mm}
    \caption{\footnotesize Text to Code Generation}
    \label{fig:concode-few-shot}
    \end{subfigure}%
    \begin{subfigure}{0.25\textwidth}
    \centering
    \includegraphics[width=\linewidth]{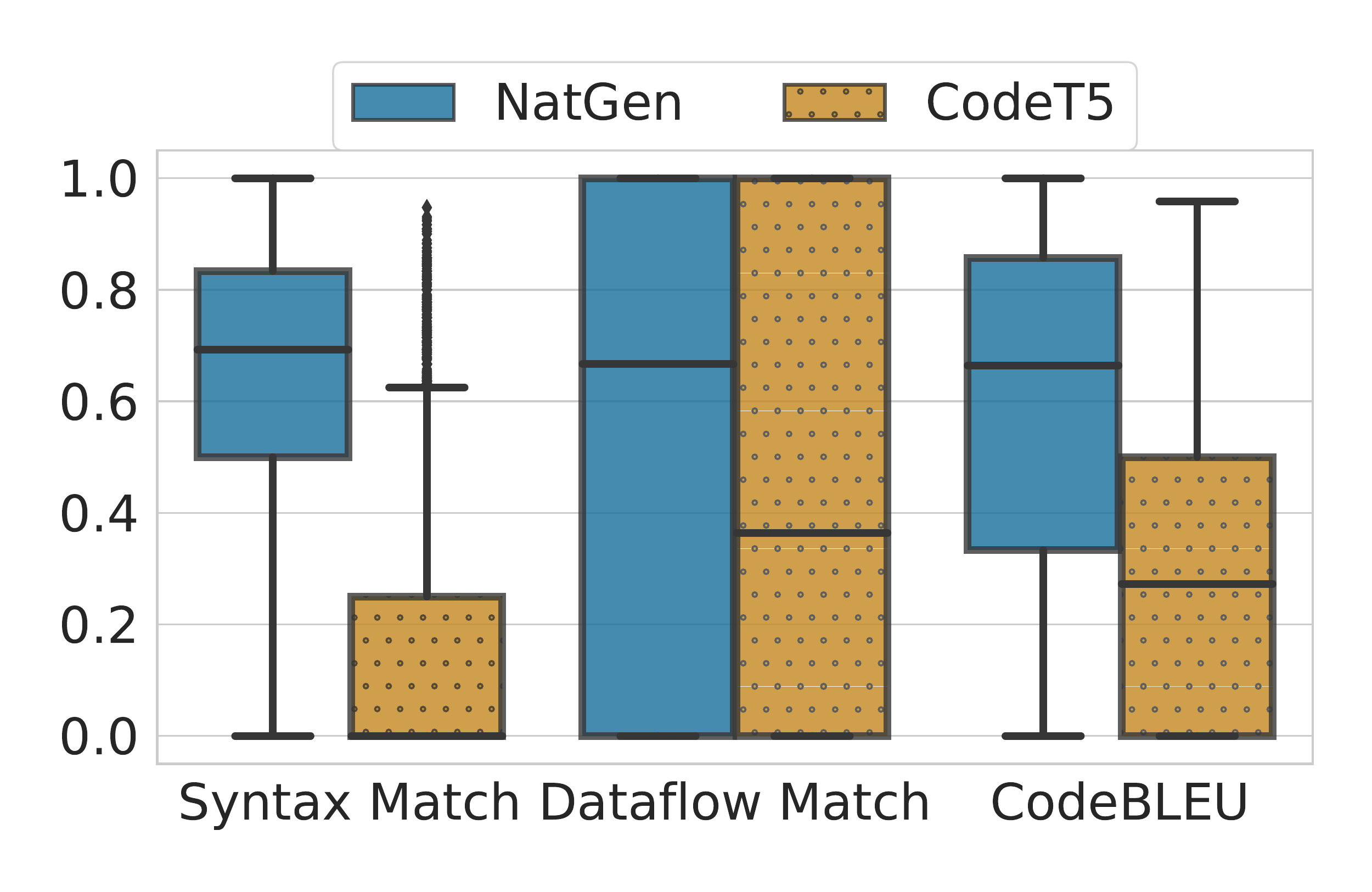}
    \vspace{-8mm}
    \caption{\footnotesize Bug Fix (small, multimodal).}
    \label{fig:refine-few-shot}
    \end{subfigure}
    \vspace{-0.2cm}
    \caption{Few shot Learning evaluation of \tool. In each case, the pre-trained model is fine-tuned on 200 training examples for 10 epoch and the result is on the full test set.}
    \label{fig:few-shot-boxplot}
\end{figure*}
\begin{figure*}[!tbh]
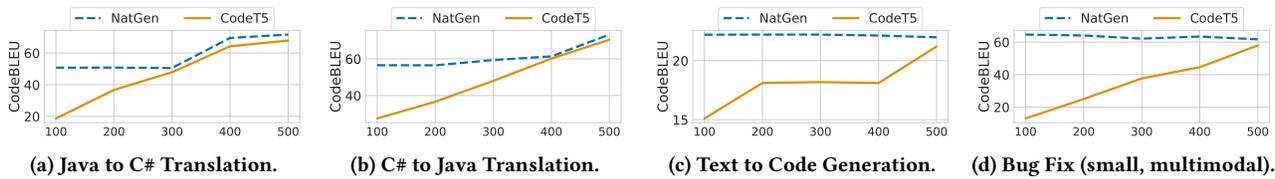

    \centering
    \input{figures/rq3/few-shot-progress/few-shot-progress-java-cs}%
    \input{figures/rq3/few-shot-progress/few-shot-progress-cs-java}
    \input{figures/rq3/few-shot-progress/few-shot-progress-concode}%
    \input{figures/rq3/few-shot-progress/few-shot-progress-refine-commit-small}
    \vspace{-3mm}
    \caption{\tool's results on different tasks with Few shot settings.
X-axis shows number of training examples.}
    \label{fig:few-shot-progress}
\end{figure*}

\noindent\textbf{Bug Fix.} 
Similar to MODIT, we evaluate the top-1 accuracy of the generated fixed code. We also evaluate uni-modal settings, where the fix description is unavailable, and multi-modal settings, where we have access to the fix description.
\Cref{tab:code_refinement} shows the results of Bug Fix. 
For the BugFix$_{small}$ dataset, \tool outperforms both CodeT5 and MODIT in both unimodal and multi-modal settings. 
For For the BugFix$_{medium}$ dataset, \tool performs better than CodeT5 and MODIT in unimodal setting and slightly worse than CodeT5 in the multi-modal setting.

\RS{3}{
\tool performs better than most of the existing baselines. 
\tool's improvement in Syntax match and Dataflow match signifies \tool's ability to generate code syntactically and semantically closer to target code.}




\RQrepeat{4}{\rqd}
\label{sec:rq4}

\paragraph{\underline{Motivation.}} Learning to generate code usually
requires a large amount of annotated training data. 
A lot of time and effort goes into curating high-quality training data~\cite{ahmad2021avatar, lachaux2020unsupervised}. Unsupervised pre-training endows machine learning models with necessary domain knowledge about the task~\cite{erhan2010does}. In practice, this knowledge appears to transfer across multiple tasks. Such pre-training reduces the effort to learn each different task. We therefore study the effectiveness of \tool's domain knowledge about source code syntax and semantics. In particular, we {\em stress test} whether the knowledge \tool learned during pre-training is useful for downstream tasks, by limiting
available task-specific training data. 

\paragraph{\underline{Experimental Setup.}} We evaluate \tool's over 
several data-limited tasks: {\em Text to Code generation}, {\em Code Translation}, and {\em Bug Fix}. We consider two different settings. First, we consider zero-shot~\cite{romera2015embarrassingly, xian2018zero} evaluation. Here we evaluate different pre-trained models {\em without} any task-specific training. Naturally, we don't see good performance in this setting. Nevertheless, this stress-test measures the code generation ability of models. Second, we try few-shot learning~\cite{wang2020generalizing, ravi2016optimization, sun2019meta}. We randomly choose a few training examples for each task and fine-tune the pre-trained models on those examples, and evaluate their performance. We gradually increase the number of training examples over several few-shot settings. 

\paragraph{\underline{Results.}} \Cref{fig:zero-shot-boxplot} shows the \tool's and CodeT5's zero-shot performance. Lacking task-specific training, we can see here how much transferable knowledge each model learned just during pre-training. There are large differences in all the tasks between \tool and CodeT5 across Syntax Match and Dataflow Match. It signifies \tool learns to generate both syntactically and semantically correct code during pre-training, which CodeT5 rarely can do. 
\Cref{fig:few-shot-boxplot} shows the performance of \tool and CodeT5 when trained on 200 training examples. \tool also has an advantage over CodeT5 here. 

We note a larger performance gap in the Translation tasks (\Cref{fig:java-cs-zero-shot} \& \ref{fig:cs-java-zero-shot}) and Bug Fix (\Cref{fig:refine-zero-shot}) tasks, compared to Text to Code Generation task (\Cref{fig:concode-zero-shot}) in both the zero-shot and the few shot (\Cref{fig:few-shot-boxplot}) experiments. We conjecture that such discrepancy is the artifact of the nature of the tasks. The cross-lingual alignment between NL and Java code is the key factor in generating text to code. In contrast, both the input and output are the programming language in the translation and bug fix task. Thus, we hypothesize that \tool leverages its shared knowledge across different programming languages learned during the pre-training. 

We further stress test \tool's with few-shot learning; we gradually increased the number of training examples and trained both CodeT5 and \tool. \Cref{fig:few-shot-progress} shows the performance progress as the number of training examples increase. 
For all four tasks, \tool significantly improves over CodeT5 when the number of training examples is minimal. With increasing training examples, the performance gap gradually decreases. Arguably, with enough {\em labeled} data and enough resources, all high-capacity models will get better at generating source code. Nevertheless, we learn two critical lessons from \tool's better performance in zero-shot and few-shot learning. First, \tool's better performance across all tasks suggests  that that the coding knowledge it learns from the naturalization task is more generic and transferable. Second, for any pre-trained model to be effective in code generation, especially in a limited training data scenario, the pre-training should explicitly teach the model how to write code. Otherwise, we hypothesize that a big chunk of fine-tuning resources will be spent on the models' learning to write code. 

\RS{4}{
\tool is very effective in source code generative tasks when minimal training resource is available. Since \tool explicitly learns to generate code during pre-training, it can avoid learning such during fine-tuning saving fine-tuning resource. 
}

\section{Limitations \& Threats}
\label{sec:discussion}

\paragraph{\textbf{Bias introduced by `de-naturalizing' transformations.}}
In \Cref{sec:semantic_denoising}, we described our six transformations to ``de-naturalize" source code. The \tool model learns to revert one transformation at a time. In fact, we found empirically that, when given code with more than one `de-naturalization' transformation applied, the model reverses only one of them. There is thus a threat our limited
application of de-naturalization limits the ability of 
our \tool. 
Regardless, we consider \tool as a proof-of-concept and the first work towards teaching a model to write natural code. We leave the investigation more natural code patterns and their effect on code generation as a potential future work.

\begin{table}[!htb]
    \small
    \caption{\tool's performance in Code summarization}
        \label{tab:code_summary}
    \centering
    \resizebox{\linewidth}{!}
    {
    \begin{tabular}{c|cccccc|c}
    \hlineB{2}
    \textbf{Approach} & \textbf{Go}&\textbf{Java} & \textbf{JS} & \textbf{Python} & \textbf{Php} & \textbf{Ruby} & \textbf{Overall} \bigstrut\\
    \hlineB{2}
    PLBART & 18.91 & 18.45 & 15.56 & 19.30 & 23.58 & 14.11 & 18.32 \bigstrut \\
    CodeT5 & \textbf{19.56} & 20.31 &  \textbf{16.16} & 20.01 &\textbf{ 26.03} & 15.24 & 19.55 \bigstrut\\ 
    \tool & 19.43 & \textbf{20.38} & 16.00 & \textbf{20.09} & 26.00 & \textbf{15.38} & 19.55 \bigstrut\\ 
    \hlineB{2}
    \end{tabular}
    }

\end{table}

\paragraph{\textbf{Knowledge retention from CodeT5.}} 
As mentioned in \Cref{sec:pretrain_exp}, we start \tool's pre-training from CodeT5-base model~\cite{wang2021codet5}. Starting further pre-training from an existing pre-trained checkpoint is very common in large-scale pre-training. For instance, GraphCodeBERT~\cite{guo2020graphcodebert}  is pre-trained based on CodeBERT~\cite{feng2020codebert} model, which was pre-trained based on RoBERTa~\cite{liu2019roberta} model. Both the Open AI-CodeX~\cite{chen2021evaluating} and Github Copilot~\cite{copilot} models are further pre-trained in OpenAI-GPT3~\cite{brown2020language}. Nevertheless, when we further train a pre-trained model on different tasks, it is subject to ``{\em catastrophic forgetting}''~\cite{kirkpatrick2017overcoming} of the knowledge learned in the base model. In order to test whether \tool is forgetting CodeT5's knowledge about natural language generation, we also evaluate \tool for Code summarization. Here the input is source code, and the output is Natural language. After fine-tuning \tool's overall BLEU in 19.547 while CodeT5's was 19.551, suggesting that \tool mostly retains CodeT5's capacity to generate NL (see \Cref{tab:code_summary} for detailed results). 

\paragraph{\textbf{Fair Comparison with CodeT5.}} 
We initialize \tool with pre-trained checkpoint from CodeT5 (already pre-trained 75K steps with their objective) and train \tool for 25K steps with `natural-code' writing objective. A skeptic reader would want to know what happens when we pre-train CodeT5 for 25K more steps with their training objective. We argue that since the pre-training objective does not explicitly account for generating code (See section 3.2 of CodeT5's original paper), further training with the CodeT5 objective does not necessarily increase its code generation capacity. We do acknowledge CodeT5's ability to understand and reason about {\em input}. Since the pre-training large model is extremely expensive~(\S\ref{sec:experiemnts})\footnote{CodeT5 was pre-trained on 16 NVIDIA A100s, with 40G memory each, for 12 days! One might reasonably assume it was already
well-trained on the original objective}; we leverage such knowledge by initializing \tool from CodeT5's publicly available pre-trained model. Moreover, CodeT5 release neither their code for pre-training (only for fine-tuning), nor any earlier or later checkpoints for us to carry out further investigation. 

\paragraph{\textbf{``Naturalization'' with program-analysis.}}
\tool is a prototype of a generative pre-trained model with ``Naturalization'' task, trained to revert six classes of de-naturalization transformations (see 
\Cref{fig:transformations}). However, perfect performance \wrt these transformation is {\bf not} the main objective of this research. Tools to accomplish ``naturalization" could surely be built using traditional refactoring methods;
however, our goal is to train \tool so that it learns to generate natural code with the help of this ``Naturalization'' task.

\paragraph{\textbf{\tool as ``Code-Refactoring'' tool.}}
\tool suggests the promise of neural transformers to 
build meaning-preserving code-refactoring tools. However,
to realize a more accurate and powerful neural re-factoring
tool,  more training data, with a larger variety of transformations, would be required. We leave this as future work.



\section{Related Works}
\label{sec:related}
The approach of pre-training large Transformers without human labels started in NLP 
domain with BERT~\cite{devlin2018bert}, which introduces two pre-training objectives (i.e., Mask Language Modeling and Next Sentence Prediction). Later, Liu et al. show that RoBERTa~\cite{liu2019roberta} outperforms BERT only using Mask Language Modeling (MLM) with new training strategies and hyper-parameter tuning. MLM is a self-supervised task that the model randomly masks or modifies a certain number of tokens and tries to recover them. 

Following the success of the pre-trained model in the NLP domain, researchers applied these models to code related tasks. 
CodeBERT is one of the earliest that was specially trained for code and relevant natural language descriptions. It is pre-trained with two objectives (i.e., MLM and Replaced Token Detection~\cite{clark2020electra}) and demonstrated pre-training's effectiveness for code. Later, an architecturally equivalent model, GraphCodeBERT, was introduced; it improved over CodeBERT on most tasks by incorporating data-flow information. 

Though CodeBERT~\cite{feng2020codebert} \& GraphCodeBERT~\cite{guo2020graphcodebert}, DietCodeBERT~\cite{zhang2022diet} do well at code understanding tasks, these models are not as good at generative tasks. Both models are encoder-only and have to start with an untrained decoder in fine-tuning for generative tasks, such as code repair, code generation, code summarization, and code translation. To address this limitation, Ahmad et al. introduced PLBART~\cite{ahmad2021unified}, pre-trained as a generative denoising autoencoder. A specific set of noises is introduced to code and relevant natural language description and used as the input to the model. The model's objective is to encode the noisy input in the encoder and generate noise-free code or text in the decoder. PLBART (builds on BART~\cite{lewis2019bart}) outperforms both CodeBERT~\cite{feng2020codebert} and GraphCodeBERT~\cite{guo2020graphcodebert} on both understanding and generative tasks with a pre-trained encoder and decoder~\cite{ahmad2021unified}. DOBF~\cite{roziere2021dobf} uses de-obfuscation (recovering variable names) as their pre-training task; however, rather than generating code, they just generate a dictionary
of recovered names. 

CodeT5~\cite{wang2021codet5} (based T5~\cite{raffel2019exploring}) is the latest denoising model. CodeT5 uses the developer-assigned identifiers in code, adding two code-specific pre-training objectives to the original T5, identifier tagging and masked identifier prediction. CodeT5 is an encoder-decoder model and excels at both understanding and generative tasks compared to other models. Similar to CodeT5, \cite{phan2021cotext,mastropaolo2021studying} are also built based on T5 architecture and perform reasonably well in the different downstream tasks. \tool has a similar architecture to CodeT5; but rather than CodeT5's pre-training objectives, we ``de-naturalize" code, using the formal channel of code to inject meaning-preserving transforms, and then force \tool to recreate, the original, ``natural" code. Rewriting semantically equivalent code requires semantic understanding, and that can be applied to code only because of its dual-channel nature. Our evaluation shows that rewriting semantically equivalent programs in the pre-training stage results in 
performance gains in at least three popular Software Engineering tasks.

\section{Conclusion}
\label{sec:conclusion}
We introduce the ``Code-Naturalization'' pre-training objective for generative models of code. 
As proof-of-concept  we pre-trained our \tool to write `natural' source code from `un-natural' counterpart. 
With this pre-training, \tool learns to write code syntactically and semantically closer to developers' written code. 
We ``de-naturalize''  existing developers'  code, using six kinds of ``semantic-preserving'' transformations. 
We further fine-tune the \tool on different variations of three downstream tasks that require code generation. 
\tool achieves state-of-the-art performance in these downstream tasks, and \tool's generated code are syntactically and semantically closer to the target code. 
Our pre-training on the `naturalizing' task is especially effective in resource-constrained setting \ie zero-shot, and few-shot transfer learning. 


\balance
\bibliographystyle{ACM-Reference-Format}
\bibliography{main}


\begin{thebibliography}{67}


\ifx \showCODEN    \undefined \def \showCODEN     #1{\unskip}     \fi
\ifx \showDOI      \undefined \def \showDOI       #1{#1}\fi
\ifx \showISBNx    \undefined \def \showISBNx     #1{\unskip}     \fi
\ifx \showISBNxiii \undefined \def \showISBNxiii  #1{\unskip}     \fi
\ifx \showISSN     \undefined \def \showISSN      #1{\unskip}     \fi
\ifx \showLCCN     \undefined \def \showLCCN      #1{\unskip}     \fi
\ifx \shownote     \undefined \def \shownote      #1{#1}          \fi
\ifx \showarticletitle \undefined \def \showarticletitle #1{#1}   \fi
\ifx \showURL      \undefined \def \showURL       {\relax}        \fi
\providecommand\bibfield[2]{#2}
\providecommand\bibinfo[2]{#2}
\providecommand\natexlab[1]{#1}
\providecommand\showeprint[2][]{arXiv:#2}

\bibitem[\protect\citeauthoryear{??}{cop}{[n.d.]}]%
        {copilot}
 \bibinfo{year}{[n.d.]}\natexlab{}.
\newblock \bibinfo{title}{\href{{{https://copilot.github.com/}}}{Github
  Copilot}, source = {\url{https://copilot.github.com/}},}.
\newblock
\newblock


\bibitem[\protect\citeauthoryear{??}{cod}{[n.d.]}]%
        {codebleu_calculator}
 \bibinfo{year}{[n.d.]}\natexlab{}.
\newblock
  \bibinfo{title}{\href{{https://github.com/microsoft/CodeXGLUE/tree/main/Code-Code/code-to-code-trans/evaluator/CodeBLEU}}{Microsoft's
  tool for CodeBLEU calculation}, source =
  {\url{https://github.com/microsoft/CodeXGLUE/tree/main/Code-Code/code-to-code-trans/evaluator/CodeBLEU}},}.
\newblock
\newblock


\bibitem[\protect\citeauthoryear{??}{cxg}{[n.d.]}]%
        {cxg_leaderboard}
 \bibinfo{year}{[n.d.]}\natexlab{}.
\newblock
  \bibinfo{title}{\href{https://microsoft.github.io/CodeXGLUE/}{CodeXGLUE
  Leaderboard.}}
\newblock
\newblock


\bibitem[\protect\citeauthoryear{Ahmad, Chakraborty, Ray, and Chang}{Ahmad
  et~al\mbox{.}}{2020}]%
        {ahmad2020summarization}
\bibfield{author}{\bibinfo{person}{Wasi~Uddin Ahmad}, \bibinfo{person}{Saikat
  Chakraborty}, \bibinfo{person}{Baishakhi Ray}, {and} \bibinfo{person}{Kai-Wei
  Chang}.} \bibinfo{year}{2020}\natexlab{}.
\newblock \showarticletitle{A Transformer-based Approach for Source Code
  Summarization}. In \bibinfo{booktitle}{\emph{Proceedings of the 58th Annual
  Meeting of the Association for Computational Linguistics (ACL)}}.
\newblock


\bibitem[\protect\citeauthoryear{Ahmad, Chakraborty, Ray, and Chang}{Ahmad
  et~al\mbox{.}}{2021a}]%
        {ahmad2021unified}
\bibfield{author}{\bibinfo{person}{Wasi~Uddin Ahmad}, \bibinfo{person}{Saikat
  Chakraborty}, \bibinfo{person}{Baishakhi Ray}, {and} \bibinfo{person}{Kai-Wei
  Chang}.} \bibinfo{year}{2021}\natexlab{a}.
\newblock \showarticletitle{Unified Pre-training for Program Understanding and
  Generation}. In \bibinfo{booktitle}{\emph{2021 Annual Conference of the North
  American Chapter of the Association for Computational Linguistics (NAACL)}}.
\newblock


\bibitem[\protect\citeauthoryear{Ahmad, Tushar, Chakraborty, and Chang}{Ahmad
  et~al\mbox{.}}{2021b}]%
        {ahmad2021avatar}
\bibfield{author}{\bibinfo{person}{Wasi~Uddin Ahmad},
  \bibinfo{person}{Md~Golam~Rahman Tushar}, \bibinfo{person}{Saikat
  Chakraborty}, {and} \bibinfo{person}{Kai-Wei Chang}.}
  \bibinfo{year}{2021}\natexlab{b}.
\newblock \bibinfo{title}{AVATAR: A Parallel Corpus for Java-Python Program
  Translation}.
\newblock
\newblock
\showeprint[arxiv]{2108.11590}~[cs.SE]


\bibitem[\protect\citeauthoryear{Allamanis}{Allamanis}{2019}]%
        {allamanis2019adverse}
\bibfield{author}{\bibinfo{person}{Miltiadis Allamanis}.}
  \bibinfo{year}{2019}\natexlab{}.
\newblock \showarticletitle{The adverse effects of code duplication in machine
  learning models of code}. In \bibinfo{booktitle}{\emph{Proceedings of the
  2019 ACM SIGPLAN International Symposium on New Ideas, New Paradigms, and
  Reflections on Programming and Software}}. \bibinfo{pages}{143--153}.
\newblock


\bibitem[\protect\citeauthoryear{Allamanis, Barr, Bird, and Sutton}{Allamanis
  et~al\mbox{.}}{2015}]%
        {allamanis2015suggesting}
\bibfield{author}{\bibinfo{person}{Miltiadis Allamanis},
  \bibinfo{person}{Earl~T Barr}, \bibinfo{person}{Christian Bird}, {and}
  \bibinfo{person}{Charles Sutton}.} \bibinfo{year}{2015}\natexlab{}.
\newblock \showarticletitle{Suggesting accurate method and class names}. In
  \bibinfo{booktitle}{\emph{Proceedings of the 2015 10th Joint Meeting on
  Foundations of Software Engineering}}. \bibinfo{pages}{38--49}.
\newblock


\bibitem[\protect\citeauthoryear{Allamanis, Barr, Devanbu, and
  Sutton}{Allamanis et~al\mbox{.}}{2018}]%
        {allamanis2018survey}
\bibfield{author}{\bibinfo{person}{Miltiadis Allamanis},
  \bibinfo{person}{Earl~T Barr}, \bibinfo{person}{Premkumar Devanbu}, {and}
  \bibinfo{person}{Charles Sutton}.} \bibinfo{year}{2018}\natexlab{}.
\newblock \showarticletitle{A survey of machine learning for big code and
  naturalness}.
\newblock \bibinfo{journal}{\emph{ACM Computing Surveys (CSUR)}}
  \bibinfo{volume}{51}, \bibinfo{number}{4} (\bibinfo{year}{2018}),
  \bibinfo{pages}{1--37}.
\newblock


\bibitem[\protect\citeauthoryear{Allamanis, Barr, Ducousso, and Gao}{Allamanis
  et~al\mbox{.}}{2020}]%
        {allamanis2020typilus}
\bibfield{author}{\bibinfo{person}{Miltiadis Allamanis},
  \bibinfo{person}{Earl~T Barr}, \bibinfo{person}{Soline Ducousso}, {and}
  \bibinfo{person}{Zheng Gao}.} \bibinfo{year}{2020}\natexlab{}.
\newblock \showarticletitle{Typilus: Neural type hints}. In
  \bibinfo{booktitle}{\emph{Proceedings of the 41st acm sigplan conference on
  programming language design and implementation}}. \bibinfo{pages}{91--105}.
\newblock


\bibitem[\protect\citeauthoryear{Allamanis, Brockschmidt, and
  Khademi}{Allamanis et~al\mbox{.}}{2017}]%
        {allamanis2017learning}
\bibfield{author}{\bibinfo{person}{Miltiadis Allamanis}, \bibinfo{person}{Marc
  Brockschmidt}, {and} \bibinfo{person}{Mahmoud Khademi}.}
  \bibinfo{year}{2017}\natexlab{}.
\newblock \showarticletitle{Learning to represent programs with graphs}.
\newblock \bibinfo{journal}{\emph{arXiv preprint arXiv:1711.00740}}
  (\bibinfo{year}{2017}).
\newblock


\bibitem[\protect\citeauthoryear{Amodio, Chaudhuri, and Reps}{Amodio
  et~al\mbox{.}}{2017}]%
        {amodio2017neural}
\bibfield{author}{\bibinfo{person}{Matthew Amodio}, \bibinfo{person}{Swarat
  Chaudhuri}, {and} \bibinfo{person}{Thomas~W Reps}.}
  \bibinfo{year}{2017}\natexlab{}.
\newblock \showarticletitle{Neural attribute machines for program generation}.
\newblock \bibinfo{journal}{\emph{arXiv preprint arXiv:1705.09231}}
  (\bibinfo{year}{2017}).
\newblock


\bibitem[\protect\citeauthoryear{Brown, Mann, Ryder, Subbiah, Kaplan, Dhariwal,
  Neelakantan, Shyam, Sastry, Askell, Agarwal, Herbert-Voss, Krueger, Henighan,
  Child, Ramesh, Ziegler, Wu, Winter, Hesse, Chen, Sigler, Litwin, Gray, Chess,
  Clark, Berner, McCandlish, Radford, Sutskever, and Amodei}{Brown
  et~al\mbox{.}}{2020}]%
        {brown2020language}
\bibfield{author}{\bibinfo{person}{Tom~B. Brown}, \bibinfo{person}{Benjamin
  Mann}, \bibinfo{person}{Nick Ryder}, \bibinfo{person}{Melanie Subbiah},
  \bibinfo{person}{Jared Kaplan}, \bibinfo{person}{Prafulla Dhariwal},
  \bibinfo{person}{Arvind Neelakantan}, \bibinfo{person}{Pranav Shyam},
  \bibinfo{person}{Girish Sastry}, \bibinfo{person}{Amanda Askell},
  \bibinfo{person}{Sandhini Agarwal}, \bibinfo{person}{Ariel Herbert-Voss},
  \bibinfo{person}{Gretchen Krueger}, \bibinfo{person}{Tom Henighan},
  \bibinfo{person}{Rewon Child}, \bibinfo{person}{Aditya Ramesh},
  \bibinfo{person}{Daniel~M. Ziegler}, \bibinfo{person}{Jeffrey Wu},
  \bibinfo{person}{Clemens Winter}, \bibinfo{person}{Christopher Hesse},
  \bibinfo{person}{Mark Chen}, \bibinfo{person}{Eric Sigler},
  \bibinfo{person}{Mateusz Litwin}, \bibinfo{person}{Scott Gray},
  \bibinfo{person}{Benjamin Chess}, \bibinfo{person}{Jack Clark},
  \bibinfo{person}{Christopher Berner}, \bibinfo{person}{Sam McCandlish},
  \bibinfo{person}{Alec Radford}, \bibinfo{person}{Ilya Sutskever}, {and}
  \bibinfo{person}{Dario Amodei}.} \bibinfo{year}{2020}\natexlab{}.
\newblock \bibinfo{title}{Language Models are Few-Shot Learners}.
\newblock
\newblock
\showeprint[arxiv]{2005.14165}~[cs.CL]


\bibitem[\protect\citeauthoryear{Casalnuovo, Barr, Dash, Devanbu, and
  Morgan}{Casalnuovo et~al\mbox{.}}{2020a}]%
        {casalnuovo2020theory}
\bibfield{author}{\bibinfo{person}{Casey Casalnuovo}, \bibinfo{person}{Earl~T
  Barr}, \bibinfo{person}{Santanu~Kumar Dash}, \bibinfo{person}{Prem Devanbu},
  {and} \bibinfo{person}{Emily Morgan}.} \bibinfo{year}{2020}\natexlab{a}.
\newblock \showarticletitle{A theory of dual channel constraints}. In
  \bibinfo{booktitle}{\emph{2020 IEEE/ACM 42nd International Conference on
  Software Engineering: New Ideas and Emerging Results (ICSE-NIER)}}. IEEE,
  \bibinfo{pages}{25--28}.
\newblock


\bibitem[\protect\citeauthoryear{Casalnuovo, Lee, Wang, Devanbu, and
  Morgan}{Casalnuovo et~al\mbox{.}}{2020b}]%
        {casalnuovo2020programmers}
\bibfield{author}{\bibinfo{person}{Casey Casalnuovo}, \bibinfo{person}{Kevin
  Lee}, \bibinfo{person}{Hulin Wang}, \bibinfo{person}{Prem Devanbu}, {and}
  \bibinfo{person}{Emily Morgan}.} \bibinfo{year}{2020}\natexlab{b}.
\newblock \showarticletitle{Do programmers prefer predictable expressions in
  code?}
\newblock \bibinfo{journal}{\emph{Cognitive science}} \bibinfo{volume}{44},
  \bibinfo{number}{12} (\bibinfo{year}{2020}), \bibinfo{pages}{e12921}.
\newblock


\bibitem[\protect\citeauthoryear{Casalnuovo, Morgan, and Devanbu}{Casalnuovo
  et~al\mbox{.}}{2020c}]%
        {casalnuovo2020does}
\bibfield{author}{\bibinfo{person}{Casey Casalnuovo}, \bibinfo{person}{E
  Morgan}, {and} \bibinfo{person}{P Devanbu}.}
  \bibinfo{year}{2020}\natexlab{c}.
\newblock \showarticletitle{Does surprisal predict code comprehension
  difficulty}. In \bibinfo{booktitle}{\emph{Proceedings of the 42nd Annual
  Meeting of the Cognitive Science Society}}. Cognitive Science Society
  Toronto, Canada.
\newblock


\bibitem[\protect\citeauthoryear{Chakraborty, Ding, Allamanis, and
  Ray}{Chakraborty et~al\mbox{.}}{2020}]%
        {chakraborty2020codit}
\bibfield{author}{\bibinfo{person}{Saikat Chakraborty},
  \bibinfo{person}{Yangruibo Ding}, \bibinfo{person}{Miltiadis Allamanis},
  {and} \bibinfo{person}{Baishakhi Ray}.} \bibinfo{year}{2020}\natexlab{}.
\newblock \showarticletitle{CODIT: Code Editing with Tree-Based Neural Models}.
\newblock \bibinfo{journal}{\emph{IEEE Transactions on Software Engineering}}
  \bibinfo{volume}{1} (\bibinfo{year}{2020}), \bibinfo{pages}{1--1}.
\newblock


\bibitem[\protect\citeauthoryear{Chakraborty and Ray}{Chakraborty and
  Ray}{2021}]%
        {chakraborty2021on}
\bibfield{author}{\bibinfo{person}{Saikat Chakraborty} {and}
  \bibinfo{person}{Baishakhi Ray}.} \bibinfo{year}{2021}\natexlab{}.
\newblock \showarticletitle{On Multi-Modal Learning of Editing Source Code}. In
  \bibinfo{booktitle}{\emph{2021 36th IEEE/ACM International Conference on
  Automated Software Engineering (ASE)}}. \bibinfo{pages}{443--455}.
\newblock
\urldef\tempurl%
\url{https://doi.org/10.1109/ASE51524.2021.9678559}
\showDOI{\tempurl}


\bibitem[\protect\citeauthoryear{Chen, Tworek, Jun, Yuan, de~Oliveira~Pinto,
  Kaplan, Edwards, Burda, Joseph, Brockman, Ray, Puri, Krueger, Petrov, Khlaaf,
  Sastry, Mishkin, Chan, Gray, Ryder, Pavlov, Power, Kaiser, Bavarian, Winter,
  Tillet, Such, Cummings, Plappert, Chantzis, Barnes, Herbert-Voss, Guss,
  Nichol, Paino, Tezak, Tang, Babuschkin, Balaji, Jain, Saunders, Hesse, Carr,
  Leike, Achiam, Misra, Morikawa, Radford, Knight, Brundage, Murati, Mayer,
  Welinder, McGrew, Amodei, McCandlish, Sutskever, and Zaremba}{Chen
  et~al\mbox{.}}{2021}]%
        {chen2021evaluating}
\bibfield{author}{\bibinfo{person}{Mark Chen}, \bibinfo{person}{Jerry Tworek},
  \bibinfo{person}{Heewoo Jun}, \bibinfo{person}{Qiming Yuan},
  \bibinfo{person}{Henrique~Ponde de Oliveira~Pinto}, \bibinfo{person}{Jared
  Kaplan}, \bibinfo{person}{Harri Edwards}, \bibinfo{person}{Yuri Burda},
  \bibinfo{person}{Nicholas Joseph}, \bibinfo{person}{Greg Brockman},
  \bibinfo{person}{Alex Ray}, \bibinfo{person}{Raul Puri},
  \bibinfo{person}{Gretchen Krueger}, \bibinfo{person}{Michael Petrov},
  \bibinfo{person}{Heidy Khlaaf}, \bibinfo{person}{Girish Sastry},
  \bibinfo{person}{Pamela Mishkin}, \bibinfo{person}{Brooke Chan},
  \bibinfo{person}{Scott Gray}, \bibinfo{person}{Nick Ryder},
  \bibinfo{person}{Mikhail Pavlov}, \bibinfo{person}{Alethea Power},
  \bibinfo{person}{Lukasz Kaiser}, \bibinfo{person}{Mohammad Bavarian},
  \bibinfo{person}{Clemens Winter}, \bibinfo{person}{Philippe Tillet},
  \bibinfo{person}{Felipe~Petroski Such}, \bibinfo{person}{Dave Cummings},
  \bibinfo{person}{Matthias Plappert}, \bibinfo{person}{Fotios Chantzis},
  \bibinfo{person}{Elizabeth Barnes}, \bibinfo{person}{Ariel Herbert-Voss},
  \bibinfo{person}{William~Hebgen Guss}, \bibinfo{person}{Alex Nichol},
  \bibinfo{person}{Alex Paino}, \bibinfo{person}{Nikolas Tezak},
  \bibinfo{person}{Jie Tang}, \bibinfo{person}{Igor Babuschkin},
  \bibinfo{person}{Suchir Balaji}, \bibinfo{person}{Shantanu Jain},
  \bibinfo{person}{William Saunders}, \bibinfo{person}{Christopher Hesse},
  \bibinfo{person}{Andrew~N. Carr}, \bibinfo{person}{Jan Leike},
  \bibinfo{person}{Josh Achiam}, \bibinfo{person}{Vedant Misra},
  \bibinfo{person}{Evan Morikawa}, \bibinfo{person}{Alec Radford},
  \bibinfo{person}{Matthew Knight}, \bibinfo{person}{Miles Brundage},
  \bibinfo{person}{Mira Murati}, \bibinfo{person}{Katie Mayer},
  \bibinfo{person}{Peter Welinder}, \bibinfo{person}{Bob McGrew},
  \bibinfo{person}{Dario Amodei}, \bibinfo{person}{Sam McCandlish},
  \bibinfo{person}{Ilya Sutskever}, {and} \bibinfo{person}{Wojciech Zaremba}.}
  \bibinfo{year}{2021}\natexlab{}.
\newblock \bibinfo{title}{Evaluating Large Language Models Trained on Code}.
\newblock
\newblock
\showeprint[arxiv]{2107.03374}~[cs.LG]


\bibitem[\protect\citeauthoryear{Clark, Luong, Le, and Manning}{Clark
  et~al\mbox{.}}{2020}]%
        {clark2020electra}
\bibfield{author}{\bibinfo{person}{Kevin Clark}, \bibinfo{person}{Minh-Thang
  Luong}, \bibinfo{person}{Quoc~V. Le}, {and} \bibinfo{person}{Christopher~D.
  Manning}.} \bibinfo{year}{2020}\natexlab{}.
\newblock \showarticletitle{{ELECTRA}: Pre-training Text Encoders as
  Discriminators Rather Than Generators}. In
  \bibinfo{booktitle}{\emph{International Conference on Learning
  Representations}}.
\newblock
\urldef\tempurl%
\url{https://openreview.net/pdf?id=r1xMH1BtvB}
\showURL{%
\tempurl}


\bibitem[\protect\citeauthoryear{Devlin, Chang, Lee, and Toutanova}{Devlin
  et~al\mbox{.}}{[n.d.]}]%
        {devlin2018bert}
\bibfield{author}{\bibinfo{person}{Jacob Devlin}, \bibinfo{person}{Ming-Wei
  Chang}, \bibinfo{person}{Kenton Lee}, {and} \bibinfo{person}{Kristina
  Toutanova}.} \bibinfo{year}{[n.d.]}\natexlab{}.
\newblock \showarticletitle{{BERT}: Pre-training of Deep Bidirectional
  Transformers for Language Understanding}. In
  \bibinfo{booktitle}{\emph{Proceedings of the 2019 Conference of the North
  {A}merican Chapter of the Association for Computational Linguistics: Human
  Language Technologies, Volume 1 (Long and Short Papers)}}.
\newblock


\bibitem[\protect\citeauthoryear{Ding, Buratti, Pujar, Morari, Ray, and
  Chakraborty}{Ding et~al\mbox{.}}{2021}]%
        {ding2021contrastive}
\bibfield{author}{\bibinfo{person}{Yangruibo Ding}, \bibinfo{person}{Luca
  Buratti}, \bibinfo{person}{Saurabh Pujar}, \bibinfo{person}{Alessandro
  Morari}, \bibinfo{person}{Baishakhi Ray}, {and} \bibinfo{person}{Saikat
  Chakraborty}.} \bibinfo{year}{2021}\natexlab{}.
\newblock \bibinfo{title}{Contrastive Learning for Source Code with Structural
  and Functional Properties}.
\newblock
\newblock
\showeprint[arxiv]{2110.03868}~[cs.PL]


\bibitem[\protect\citeauthoryear{Erhan, Courville, Bengio, and Vincent}{Erhan
  et~al\mbox{.}}{2010}]%
        {erhan2010does}
\bibfield{author}{\bibinfo{person}{Dumitru Erhan}, \bibinfo{person}{Aaron
  Courville}, \bibinfo{person}{Yoshua Bengio}, {and} \bibinfo{person}{Pascal
  Vincent}.} \bibinfo{year}{2010}\natexlab{}.
\newblock \showarticletitle{Why does unsupervised pre-training help deep
  learning?}. In \bibinfo{booktitle}{\emph{Proceedings of the thirteenth
  international conference on artificial intelligence and statistics}}. JMLR
  Workshop and Conference Proceedings, \bibinfo{pages}{201--208}.
\newblock


\bibitem[\protect\citeauthoryear{Feng, Guo, Tang, Duan, Feng, Gong, Shou, Qin,
  Liu, Jiang, and Zhou}{Feng et~al\mbox{.}}{2020}]%
        {feng2020codebert}
\bibfield{author}{\bibinfo{person}{Zhangyin Feng}, \bibinfo{person}{Daya Guo},
  \bibinfo{person}{Duyu Tang}, \bibinfo{person}{Nan Duan},
  \bibinfo{person}{Xiaocheng Feng}, \bibinfo{person}{Ming Gong},
  \bibinfo{person}{Linjun Shou}, \bibinfo{person}{Bing Qin},
  \bibinfo{person}{Ting Liu}, \bibinfo{person}{Daxin Jiang}, {and}
  \bibinfo{person}{Ming Zhou}.} \bibinfo{year}{2020}\natexlab{}.
\newblock \showarticletitle{{C}ode{BERT}: A Pre-Trained Model for Programming
  and Natural Languages}. In \bibinfo{booktitle}{\emph{Findings of the
  Association for Computational Linguistics: EMNLP 2020}}.
  \bibinfo{pages}{1536--1547}.
\newblock


\bibitem[\protect\citeauthoryear{Fowler}{Fowler}{2018}]%
        {fowler2018refactoring}
\bibfield{author}{\bibinfo{person}{Martin Fowler}.}
  \bibinfo{year}{2018}\natexlab{}.
\newblock \bibinfo{booktitle}{\emph{Refactoring: improving the design of
  existing code}}.
\newblock \bibinfo{publisher}{Addison-Wesley Professional}.
\newblock


\bibitem[\protect\citeauthoryear{Gopstein, Fayard, Apel, and Cappos}{Gopstein
  et~al\mbox{.}}{2020}]%
        {gopstein2020thinking}
\bibfield{author}{\bibinfo{person}{Dan Gopstein}, \bibinfo{person}{Anne-Laure
  Fayard}, \bibinfo{person}{Sven Apel}, {and} \bibinfo{person}{Justin Cappos}.}
  \bibinfo{year}{2020}\natexlab{}.
\newblock \showarticletitle{Thinking aloud about confusing code: A qualitative
  investigation of program comprehension and atoms of confusion}. In
  \bibinfo{booktitle}{\emph{Proceedings of the 28th ACM Joint Meeting on
  European Software Engineering Conference and Symposium on the Foundations of
  Software Engineering}}. \bibinfo{pages}{605--616}.
\newblock


\bibitem[\protect\citeauthoryear{Gopstein, Zhou, Frankl, and Cappos}{Gopstein
  et~al\mbox{.}}{2018}]%
        {gopstein2018prevalence}
\bibfield{author}{\bibinfo{person}{Dan Gopstein},
  \bibinfo{person}{Hongwei~Henry Zhou}, \bibinfo{person}{Phyllis Frankl}, {and}
  \bibinfo{person}{Justin Cappos}.} \bibinfo{year}{2018}\natexlab{}.
\newblock \showarticletitle{Prevalence of confusing code in software projects:
  Atoms of confusion in the wild}. In \bibinfo{booktitle}{\emph{Proceedings of
  the 15th International Conference on Mining Software Repositories}}.
  \bibinfo{pages}{281--291}.
\newblock


\bibitem[\protect\citeauthoryear{Gros, Sezhiyan, Devanbu, and Yu}{Gros
  et~al\mbox{.}}{2020}]%
        {gros2020code}
\bibfield{author}{\bibinfo{person}{David Gros}, \bibinfo{person}{Hariharan
  Sezhiyan}, \bibinfo{person}{Prem Devanbu}, {and} \bibinfo{person}{Zhou Yu}.}
  \bibinfo{year}{2020}\natexlab{}.
\newblock \showarticletitle{Code to Comment “Translation”: Data, Metrics,
  Baselining \& Evaluation}. In \bibinfo{booktitle}{\emph{2020 35th IEEE/ACM
  International Conference on Automated Software Engineering (ASE)}}. IEEE,
  \bibinfo{pages}{746--757}.
\newblock


\bibitem[\protect\citeauthoryear{Guo, Ren, Lu, Feng, Tang, Liu, Zhou, Duan,
  Yin, Jiang, et~al\mbox{.}}{Guo et~al\mbox{.}}{2021}]%
        {guo2020graphcodebert}
\bibfield{author}{\bibinfo{person}{Daya Guo}, \bibinfo{person}{Shuo Ren},
  \bibinfo{person}{Shuai Lu}, \bibinfo{person}{Zhangyin Feng},
  \bibinfo{person}{Duyu Tang}, \bibinfo{person}{Shujie Liu},
  \bibinfo{person}{Long Zhou}, \bibinfo{person}{Nan Duan},
  \bibinfo{person}{Jian Yin}, \bibinfo{person}{Daxin Jiang}, {et~al\mbox{.}}}
  \bibinfo{year}{2021}\natexlab{}.
\newblock \showarticletitle{GraphCodeBERT: Pre-training Code Representations
  with Data Flow}. In \bibinfo{booktitle}{\emph{International Conference on
  Learning Representations}}.
\newblock


\bibitem[\protect\citeauthoryear{Guo, Tang, Duan, Zhou, and Yin}{Guo
  et~al\mbox{.}}{2019}]%
        {guo-etal-2019-coupling}
\bibfield{author}{\bibinfo{person}{Daya Guo}, \bibinfo{person}{Duyu Tang},
  \bibinfo{person}{Nan Duan}, \bibinfo{person}{Ming Zhou}, {and}
  \bibinfo{person}{Jian Yin}.} \bibinfo{year}{2019}\natexlab{}.
\newblock \showarticletitle{Coupling Retrieval and Meta-Learning for
  Context-Dependent Semantic Parsing}. In \bibinfo{booktitle}{\emph{Proceedings
  of the 57th Annual Meeting of the Association for Computational
  Linguistics}}. \bibinfo{publisher}{Association for Computational
  Linguistics}, \bibinfo{address}{Florence, Italy}, \bibinfo{pages}{855--866}.
\newblock
\urldef\tempurl%
\url{https://doi.org/10.18653/v1/P19-1082}
\showDOI{\tempurl}


\bibitem[\protect\citeauthoryear{Gupta, Pal, Kanade, and Shevade}{Gupta
  et~al\mbox{.}}{2017}]%
        {gupta2017deepfix}
\bibfield{author}{\bibinfo{person}{Rahul Gupta}, \bibinfo{person}{Soham Pal},
  \bibinfo{person}{Aditya Kanade}, {and} \bibinfo{person}{Shirish Shevade}.}
  \bibinfo{year}{2017}\natexlab{}.
\newblock \showarticletitle{{DeepFix}: Fixing Common {C} Language Errors by
  Deep Learning.}. In \bibinfo{booktitle}{\emph{AAAI}}.
  \bibinfo{pages}{1345--1351}.
\newblock


\bibitem[\protect\citeauthoryear{Hellendoorn and Devanbu}{Hellendoorn and
  Devanbu}{2017}]%
        {hellendoorn2017deep}
\bibfield{author}{\bibinfo{person}{Vincent~J Hellendoorn} {and}
  \bibinfo{person}{Premkumar Devanbu}.} \bibinfo{year}{2017}\natexlab{}.
\newblock \showarticletitle{Are deep neural networks the best choice for
  modeling source code?}. In \bibinfo{booktitle}{\emph{Proceedings of the 2017
  11th Joint Meeting on Foundations of Software Engineering}}. ACM,
  \bibinfo{pages}{763--773}.
\newblock


\bibitem[\protect\citeauthoryear{Hindle, Barr, Gabel, Su, and Devanbu}{Hindle
  et~al\mbox{.}}{2016}]%
        {hindle2016naturalness}
\bibfield{author}{\bibinfo{person}{Abram Hindle}, \bibinfo{person}{Earl~T
  Barr}, \bibinfo{person}{Mark Gabel}, \bibinfo{person}{Zhendong Su}, {and}
  \bibinfo{person}{Premkumar Devanbu}.} \bibinfo{year}{2016}\natexlab{}.
\newblock \showarticletitle{On the naturalness of software}.
\newblock \bibinfo{journal}{\emph{Commun. ACM}} \bibinfo{volume}{59},
  \bibinfo{number}{5} (\bibinfo{year}{2016}), \bibinfo{pages}{122--131}.
\newblock


\bibitem[\protect\citeauthoryear{Hindle, Barr, Su, Gabel, and Devanbu}{Hindle
  et~al\mbox{.}}{2012}]%
        {hindle2012naturalness}
\bibfield{author}{\bibinfo{person}{Abram Hindle}, \bibinfo{person}{Earl~T
  Barr}, \bibinfo{person}{Zhendong Su}, \bibinfo{person}{Mark Gabel}, {and}
  \bibinfo{person}{Premkumar Devanbu}.} \bibinfo{year}{2012}\natexlab{}.
\newblock \showarticletitle{On the naturalness of software}. In
  \bibinfo{booktitle}{\emph{2012 34th International Conference on Software
  Engineering (ICSE)}}. IEEE, \bibinfo{pages}{837--847}.
\newblock


\bibitem[\protect\citeauthoryear{Husain, Wu, Gazit, Allamanis, and
  Brockschmidt}{Husain et~al\mbox{.}}{2019}]%
        {husain2019codesearchnet}
\bibfield{author}{\bibinfo{person}{Hamel Husain}, \bibinfo{person}{Ho-Hsiang
  Wu}, \bibinfo{person}{Tiferet Gazit}, \bibinfo{person}{Miltiadis Allamanis},
  {and} \bibinfo{person}{Marc Brockschmidt}.} \bibinfo{year}{2019}\natexlab{}.
\newblock \showarticletitle{Codesearchnet challenge: Evaluating the state of
  semantic code search}.
\newblock \bibinfo{journal}{\emph{arXiv preprint arXiv:1909.09436}}
  (\bibinfo{year}{2019}).
\newblock
\urldef\tempurl%
\url{https://arxiv.org/abs/1909.09436}
\showURL{%
\tempurl}


\bibitem[\protect\citeauthoryear{Iyer, Konstas, Cheung, and Zettlemoyer}{Iyer
  et~al\mbox{.}}{2016}]%
        {iyer2016summarizing}
\bibfield{author}{\bibinfo{person}{Srinivasan Iyer}, \bibinfo{person}{Ioannis
  Konstas}, \bibinfo{person}{Alvin Cheung}, {and} \bibinfo{person}{Luke
  Zettlemoyer}.} \bibinfo{year}{2016}\natexlab{}.
\newblock \showarticletitle{Summarizing Source Code using a Neural Attention
  Model}. In \bibinfo{booktitle}{\emph{Proceedings of the 54th Annual Meeting
  of the Association for Computational Linguistics (Volume 1: Long Papers)}}.
  \bibinfo{pages}{2073--2083}.
\newblock
\urldef\tempurl%
\url{https://doi.org/10.18653/v1/P16-1195}
\showDOI{\tempurl}


\bibitem[\protect\citeauthoryear{Iyer, Konstas, Cheung, and Zettlemoyer}{Iyer
  et~al\mbox{.}}{2018}]%
        {iyer2018mapping}
\bibfield{author}{\bibinfo{person}{Srinivasan Iyer}, \bibinfo{person}{Ioannis
  Konstas}, \bibinfo{person}{Alvin Cheung}, {and} \bibinfo{person}{Luke
  Zettlemoyer}.} \bibinfo{year}{2018}\natexlab{}.
\newblock \showarticletitle{Mapping language to code in programmatic context}.
\newblock \bibinfo{journal}{\emph{arXiv preprint arXiv:1808.09588}}
  (\bibinfo{year}{2018}).
\newblock


\bibitem[\protect\citeauthoryear{Kirkpatrick, Pascanu, Rabinowitz, Veness,
  Desjardins, Rusu, Milan, Quan, Ramalho, Grabska-Barwinska,
  et~al\mbox{.}}{Kirkpatrick et~al\mbox{.}}{2017}]%
        {kirkpatrick2017overcoming}
\bibfield{author}{\bibinfo{person}{James Kirkpatrick}, \bibinfo{person}{Razvan
  Pascanu}, \bibinfo{person}{Neil Rabinowitz}, \bibinfo{person}{Joel Veness},
  \bibinfo{person}{Guillaume Desjardins}, \bibinfo{person}{Andrei~A Rusu},
  \bibinfo{person}{Kieran Milan}, \bibinfo{person}{John Quan},
  \bibinfo{person}{Tiago Ramalho}, \bibinfo{person}{Agnieszka
  Grabska-Barwinska}, {et~al\mbox{.}}} \bibinfo{year}{2017}\natexlab{}.
\newblock \showarticletitle{Overcoming catastrophic forgetting in neural
  networks}.
\newblock \bibinfo{journal}{\emph{Proceedings of the national academy of
  sciences}} \bibinfo{volume}{114}, \bibinfo{number}{13}
  (\bibinfo{year}{2017}), \bibinfo{pages}{3521--3526}.
\newblock


\bibitem[\protect\citeauthoryear{Lachaux, Roziere, Chanussot, and
  Lample}{Lachaux et~al\mbox{.}}{2020}]%
        {lachaux2020unsupervised}
\bibfield{author}{\bibinfo{person}{Marie-Anne Lachaux},
  \bibinfo{person}{Baptiste Roziere}, \bibinfo{person}{Lowik Chanussot}, {and}
  \bibinfo{person}{Guillaume Lample}.} \bibinfo{year}{2020}\natexlab{}.
\newblock \showarticletitle{Unsupervised Translation of Programming Languages}.
\newblock \bibinfo{journal}{\emph{arXiv preprint arXiv:2006.03511}}
  (\bibinfo{year}{2020}).
\newblock


\bibitem[\protect\citeauthoryear{Lewis, Liu, Goyal, Ghazvininejad, Mohamed,
  Levy, Stoyanov, and Zettlemoyer}{Lewis et~al\mbox{.}}{2019}]%
        {lewis2019bart}
\bibfield{author}{\bibinfo{person}{Mike Lewis}, \bibinfo{person}{Yinhan Liu},
  \bibinfo{person}{Naman Goyal}, \bibinfo{person}{Marjan Ghazvininejad},
  \bibinfo{person}{Abdelrahman Mohamed}, \bibinfo{person}{Omer Levy},
  \bibinfo{person}{Ves Stoyanov}, {and} \bibinfo{person}{Luke Zettlemoyer}.}
  \bibinfo{year}{2019}\natexlab{}.
\newblock \showarticletitle{Bart: Denoising sequence-to-sequence pre-training
  for natural language generation, translation, and comprehension}.
\newblock \bibinfo{journal}{\emph{arXiv preprint arXiv:1910.13461}}
  (\bibinfo{year}{2019}).
\newblock


\bibitem[\protect\citeauthoryear{Liu, Chen, Xie, Siow, and Liu}{Liu
  et~al\mbox{.}}{2020}]%
        {liu2020retrieval}
\bibfield{author}{\bibinfo{person}{Shangqing Liu}, \bibinfo{person}{Yu Chen},
  \bibinfo{person}{Xiaofei Xie}, \bibinfo{person}{Jingkai Siow}, {and}
  \bibinfo{person}{Yang Liu}.} \bibinfo{year}{2020}\natexlab{}.
\newblock \showarticletitle{Retrieval-augmented generation for code
  summarization via hybrid gnn}.
\newblock \bibinfo{journal}{\emph{arXiv preprint arXiv:2006.05405}}
  (\bibinfo{year}{2020}).
\newblock


\bibitem[\protect\citeauthoryear{Liu, Ott, Goyal, Du, Joshi, Chen, Levy, Lewis,
  Zettlemoyer, and Stoyanov}{Liu et~al\mbox{.}}{2019}]%
        {liu2019roberta}
\bibfield{author}{\bibinfo{person}{Yinhan Liu}, \bibinfo{person}{Myle Ott},
  \bibinfo{person}{Naman Goyal}, \bibinfo{person}{Jingfei Du},
  \bibinfo{person}{Mandar Joshi}, \bibinfo{person}{Danqi Chen},
  \bibinfo{person}{Omer Levy}, \bibinfo{person}{Mike Lewis},
  \bibinfo{person}{Luke Zettlemoyer}, {and} \bibinfo{person}{Veselin
  Stoyanov}.} \bibinfo{year}{2019}\natexlab{}.
\newblock \showarticletitle{RoBERTa: A Robustly Optimized BERT Pretraining
  Approach}.
\newblock \bibinfo{journal}{\emph{arXiv preprint arXiv:1907.11692}}
  (\bibinfo{year}{2019}).
\newblock
\urldef\tempurl%
\url{https://arxiv.org/abs/1907.11692}
\showURL{%
\tempurl}


\bibitem[\protect\citeauthoryear{Lu, Guo, Ren, Huang, Svyatkovskiy, Blanco,
  Clement, Drain, Jiang, Tang, et~al\mbox{.}}{Lu et~al\mbox{.}}{2021}]%
        {CodeXGLUE}
\bibfield{author}{\bibinfo{person}{Shuai Lu}, \bibinfo{person}{Daya Guo},
  \bibinfo{person}{Shuo Ren}, \bibinfo{person}{Junjie Huang},
  \bibinfo{person}{Alexey Svyatkovskiy}, \bibinfo{person}{Ambrosio Blanco},
  \bibinfo{person}{Colin Clement}, \bibinfo{person}{Dawn Drain},
  \bibinfo{person}{Daxin Jiang}, \bibinfo{person}{Duyu Tang}, {et~al\mbox{.}}}
  \bibinfo{year}{2021}\natexlab{}.
\newblock \showarticletitle{CodeXGLUE: A Machine Learning Benchmark Dataset for
  Code Understanding and Generation}.
\newblock \bibinfo{journal}{\emph{arXiv preprint arXiv:2102.04664}}
  (\bibinfo{year}{2021}).
\newblock
\urldef\tempurl%
\url{https://arxiv.org/abs/2102.04664}
\showURL{%
\tempurl}


\bibitem[\protect\citeauthoryear{Mastropaolo, Scalabrino, Cooper, Palacio,
  Poshyvanyk, Oliveto, and Bavota}{Mastropaolo et~al\mbox{.}}{2021}]%
        {mastropaolo2021studying}
\bibfield{author}{\bibinfo{person}{Antonio Mastropaolo},
  \bibinfo{person}{Simone Scalabrino}, \bibinfo{person}{Nathan Cooper},
  \bibinfo{person}{David~Nader Palacio}, \bibinfo{person}{Denys Poshyvanyk},
  \bibinfo{person}{Rocco Oliveto}, {and} \bibinfo{person}{Gabriele Bavota}.}
  \bibinfo{year}{2021}\natexlab{}.
\newblock \showarticletitle{Studying the usage of text-to-text transfer
  transformer to support code-related tasks}. In \bibinfo{booktitle}{\emph{2021
  IEEE/ACM 43rd International Conference on Software Engineering (ICSE)}}.
  IEEE, \bibinfo{pages}{336--347}.
\newblock


\bibitem[\protect\citeauthoryear{Niu, Li, Ng, Ge, Huang, and Luo}{Niu
  et~al\mbox{.}}{2022}]%
        {niu2022spt}
\bibfield{author}{\bibinfo{person}{Changan Niu}, \bibinfo{person}{Chuanyi Li},
  \bibinfo{person}{Vincent Ng}, \bibinfo{person}{Jidong Ge},
  \bibinfo{person}{Liguo Huang}, {and} \bibinfo{person}{Bin Luo}.}
  \bibinfo{year}{2022}\natexlab{}.
\newblock \showarticletitle{SPT-Code: Sequence-to-Sequence Pre-Training for
  Learning the Representation of Source Code}.
\newblock \bibinfo{journal}{\emph{arXiv preprint arXiv:2201.01549}}
  (\bibinfo{year}{2022}).
\newblock


\bibitem[\protect\citeauthoryear{Parvez, Ahmad, Chakraborty, Ray, and
  Chang}{Parvez et~al\mbox{.}}{2021}]%
        {parvez2021retrieval}
\bibfield{author}{\bibinfo{person}{Md~Rizwan Parvez},
  \bibinfo{person}{Wasi~Uddin Ahmad}, \bibinfo{person}{Saikat Chakraborty},
  \bibinfo{person}{Baishakhi Ray}, {and} \bibinfo{person}{Kai-Wei Chang}.}
  \bibinfo{year}{2021}\natexlab{}.
\newblock \showarticletitle{Retrieval Augmented Code Generation and
  Summarization}.
\newblock \bibinfo{journal}{\emph{arXiv preprint arXiv:2108.11601}}
  (\bibinfo{year}{2021}).
\newblock


\bibitem[\protect\citeauthoryear{Patra and Pradel}{Patra and Pradel}{2021}]%
        {patra2021semantic}
\bibfield{author}{\bibinfo{person}{Jibesh Patra} {and} \bibinfo{person}{Michael
  Pradel}.} \bibinfo{year}{2021}\natexlab{}.
\newblock \showarticletitle{Semantic bug seeding: a learning-based approach for
  creating realistic bugs}. In \bibinfo{booktitle}{\emph{Proceedings of the
  29th ACM Joint Meeting on European Software Engineering Conference and
  Symposium on the Foundations of Software Engineering}}.
  \bibinfo{pages}{906--918}.
\newblock


\bibitem[\protect\citeauthoryear{Phan, Tran, Le, Nguyen, Anibal, Peltekian, and
  Ye}{Phan et~al\mbox{.}}{2021}]%
        {phan2021cotext}
\bibfield{author}{\bibinfo{person}{Long Phan}, \bibinfo{person}{Hieu Tran},
  \bibinfo{person}{Daniel Le}, \bibinfo{person}{Hieu Nguyen},
  \bibinfo{person}{James Anibal}, \bibinfo{person}{Alec Peltekian}, {and}
  \bibinfo{person}{Yanfang Ye}.} \bibinfo{year}{2021}\natexlab{}.
\newblock \showarticletitle{CoTexT: Multi-task Learning with Code-Text
  Transformer}.
\newblock \bibinfo{journal}{\emph{arXiv preprint arXiv:2105.08645}}
  (\bibinfo{year}{2021}).
\newblock


\bibitem[\protect\citeauthoryear{Pradel and Chandra}{Pradel and
  Chandra}{2021}]%
        {pradel2021neural}
\bibfield{author}{\bibinfo{person}{Michael Pradel} {and}
  \bibinfo{person}{Satish Chandra}.} \bibinfo{year}{2021}\natexlab{}.
\newblock \showarticletitle{Neural software analysis}.
\newblock \bibinfo{journal}{\emph{Commun. ACM}} \bibinfo{volume}{65},
  \bibinfo{number}{1} (\bibinfo{year}{2021}), \bibinfo{pages}{86--96}.
\newblock


\bibitem[\protect\citeauthoryear{Radford, Wu, Child, Luan, Amodei, and
  Sutskever}{Radford et~al\mbox{.}}{2019}]%
        {radford2019language}
\bibfield{author}{\bibinfo{person}{Alec Radford}, \bibinfo{person}{Jeffrey Wu},
  \bibinfo{person}{Rewon Child}, \bibinfo{person}{David Luan},
  \bibinfo{person}{Dario Amodei}, {and} \bibinfo{person}{Ilya Sutskever}.}
  \bibinfo{year}{2019}\natexlab{}.
\newblock \showarticletitle{Language models are unsupervised multitask
  learners}.
\newblock \bibinfo{journal}{\emph{OpenAI blog}} \bibinfo{volume}{1},
  \bibinfo{number}{8} (\bibinfo{year}{2019}), \bibinfo{pages}{9}.
\newblock


\bibitem[\protect\citeauthoryear{Raffel, Shazeer, Roberts, Lee, Narang, Matena,
  Zhou, Li, and Liu}{Raffel et~al\mbox{.}}{2019}]%
        {raffel2019exploring}
\bibfield{author}{\bibinfo{person}{Colin Raffel}, \bibinfo{person}{Noam
  Shazeer}, \bibinfo{person}{Adam Roberts}, \bibinfo{person}{Katherine Lee},
  \bibinfo{person}{Sharan Narang}, \bibinfo{person}{Michael Matena},
  \bibinfo{person}{Yanqi Zhou}, \bibinfo{person}{Wei Li}, {and}
  \bibinfo{person}{Peter~J Liu}.} \bibinfo{year}{2019}\natexlab{}.
\newblock \showarticletitle{Exploring the limits of transfer learning with a
  unified text-to-text transformer}.
\newblock \bibinfo{journal}{\emph{arXiv preprint arXiv:1910.10683}}
  (\bibinfo{year}{2019}).
\newblock


\bibitem[\protect\citeauthoryear{Ravi and Larochelle}{Ravi and
  Larochelle}{2016}]%
        {ravi2016optimization}
\bibfield{author}{\bibinfo{person}{Sachin Ravi} {and} \bibinfo{person}{Hugo
  Larochelle}.} \bibinfo{year}{2016}\natexlab{}.
\newblock \showarticletitle{Optimization as a model for few-shot learning}.
\newblock  (\bibinfo{year}{2016}).
\newblock


\bibitem[\protect\citeauthoryear{Ray, Hellendoorn, Godhane, Tu, Bacchelli, and
  Devanbu}{Ray et~al\mbox{.}}{2016}]%
        {ray2016naturalness}
\bibfield{author}{\bibinfo{person}{Baishakhi Ray}, \bibinfo{person}{Vincent
  Hellendoorn}, \bibinfo{person}{Saheel Godhane}, \bibinfo{person}{Zhaopeng
  Tu}, \bibinfo{person}{Alberto Bacchelli}, {and} \bibinfo{person}{Premkumar
  Devanbu}.} \bibinfo{year}{2016}\natexlab{}.
\newblock \showarticletitle{On the" naturalness" of buggy code}. In
  \bibinfo{booktitle}{\emph{2016 IEEE/ACM 38th International Conference on
  Software Engineering (ICSE)}}. IEEE, \bibinfo{pages}{428--439}.
\newblock


\bibitem[\protect\citeauthoryear{Raychev, Vechev, and Yahav}{Raychev
  et~al\mbox{.}}{2014}]%
        {raychev2014code}
\bibfield{author}{\bibinfo{person}{Veselin Raychev}, \bibinfo{person}{Martin
  Vechev}, {and} \bibinfo{person}{Eran Yahav}.}
  \bibinfo{year}{2014}\natexlab{}.
\newblock \showarticletitle{Code completion with statistical language models}.
  In \bibinfo{booktitle}{\emph{Acm Sigplan Notices}},
  Vol.~\bibinfo{volume}{49}. ACM, \bibinfo{pages}{419--428}.
\newblock


\bibitem[\protect\citeauthoryear{Ren, Guo, Lu, Zhou, Liu, Tang, Zhou, Blanco,
  and Ma}{Ren et~al\mbox{.}}{2020}]%
        {ren2020codebleu}
\bibfield{author}{\bibinfo{person}{Shuo Ren}, \bibinfo{person}{Daya Guo},
  \bibinfo{person}{Shuai Lu}, \bibinfo{person}{Long Zhou},
  \bibinfo{person}{Shujie Liu}, \bibinfo{person}{Duyu Tang},
  \bibinfo{person}{Ming Zhou}, \bibinfo{person}{Ambrosio Blanco}, {and}
  \bibinfo{person}{Shuai Ma}.} \bibinfo{year}{2020}\natexlab{}.
\newblock \showarticletitle{CodeBLEU: a Method for Automatic Evaluation of Code
  Synthesis}.
\newblock \bibinfo{journal}{\emph{arXiv preprint arXiv:2009.10297}}
  (\bibinfo{year}{2020}).
\newblock
\urldef\tempurl%
\url{https://arxiv.org/abs/2009.10297}
\showURL{%
\tempurl}


\bibitem[\protect\citeauthoryear{Romera-Paredes and Torr}{Romera-Paredes and
  Torr}{2015}]%
        {romera2015embarrassingly}
\bibfield{author}{\bibinfo{person}{Bernardino Romera-Paredes} {and}
  \bibinfo{person}{Philip Torr}.} \bibinfo{year}{2015}\natexlab{}.
\newblock \showarticletitle{An embarrassingly simple approach to zero-shot
  learning}. In \bibinfo{booktitle}{\emph{International conference on machine
  learning}}. PMLR, \bibinfo{pages}{2152--2161}.
\newblock


\bibitem[\protect\citeauthoryear{Roziere, Lachaux, Szafraniec, and
  Lample}{Roziere et~al\mbox{.}}{2021}]%
        {roziere2021dobf}
\bibfield{author}{\bibinfo{person}{Baptiste Roziere},
  \bibinfo{person}{Marie-Anne Lachaux}, \bibinfo{person}{Marc Szafraniec},
  {and} \bibinfo{person}{Guillaume Lample}.} \bibinfo{year}{2021}\natexlab{}.
\newblock \showarticletitle{DOBF: A deobfuscation pre-training objective for
  programming languages}.
\newblock \bibinfo{journal}{\emph{arXiv preprint arXiv:2102.07492}}
  (\bibinfo{year}{2021}).
\newblock


\bibitem[\protect\citeauthoryear{Sun, Liu, Chua, and Schiele}{Sun
  et~al\mbox{.}}{2019}]%
        {sun2019meta}
\bibfield{author}{\bibinfo{person}{Qianru Sun}, \bibinfo{person}{Yaoyao Liu},
  \bibinfo{person}{Tat-Seng Chua}, {and} \bibinfo{person}{Bernt Schiele}.}
  \bibinfo{year}{2019}\natexlab{}.
\newblock \showarticletitle{Meta-transfer learning for few-shot learning}. In
  \bibinfo{booktitle}{\emph{Proceedings of the IEEE/CVF Conference on Computer
  Vision and Pattern Recognition}}. \bibinfo{pages}{403--412}.
\newblock


\bibitem[\protect\citeauthoryear{Tufano, Pantiuchina, Watson, Bavota, and
  Poshyvanyk}{Tufano et~al\mbox{.}}{2019a}]%
        {tufano2019learning}
\bibfield{author}{\bibinfo{person}{Michele Tufano}, \bibinfo{person}{Jevgenija
  Pantiuchina}, \bibinfo{person}{Cody Watson}, \bibinfo{person}{Gabriele
  Bavota}, {and} \bibinfo{person}{Denys Poshyvanyk}.}
  \bibinfo{year}{2019}\natexlab{a}.
\newblock \showarticletitle{On Learning Meaningful Code Changes via Neural
  Machine Translation}.
\newblock \bibinfo{journal}{\emph{arXiv preprint arXiv:1901.09102}}
  (\bibinfo{year}{2019}).
\newblock


\bibitem[\protect\citeauthoryear{Tufano, Watson, Bavota, Penta, White, and
  Poshyvanyk}{Tufano et~al\mbox{.}}{2019b}]%
        {tufano2019empirical}
\bibfield{author}{\bibinfo{person}{Michele Tufano}, \bibinfo{person}{Cody
  Watson}, \bibinfo{person}{Gabriele Bavota}, \bibinfo{person}{Massimiliano~Di
  Penta}, \bibinfo{person}{Martin White}, {and} \bibinfo{person}{Denys
  Poshyvanyk}.} \bibinfo{year}{2019}\natexlab{b}.
\newblock \showarticletitle{An empirical study on learning bug-fixing patches
  in the wild via neural machine translation}.
\newblock \bibinfo{journal}{\emph{ACM Transactions on Software Engineering and
  Methodology (TOSEM)}} \bibinfo{volume}{28}, \bibinfo{number}{4}
  (\bibinfo{year}{2019}), \bibinfo{pages}{1--29}.
\newblock


\bibitem[\protect\citeauthoryear{Vasilescu, Casalnuovo, and Devanbu}{Vasilescu
  et~al\mbox{.}}{2017}]%
        {vasilescu2017recovering}
\bibfield{author}{\bibinfo{person}{Bogdan Vasilescu}, \bibinfo{person}{Casey
  Casalnuovo}, {and} \bibinfo{person}{Premkumar Devanbu}.}
  \bibinfo{year}{2017}\natexlab{}.
\newblock \showarticletitle{Recovering clear, natural identifiers from
  obfuscated JS names}. In \bibinfo{booktitle}{\emph{Proceedings of the 2017
  11th joint meeting on foundations of software engineering}}.
  \bibinfo{pages}{683--693}.
\newblock


\bibitem[\protect\citeauthoryear{Vaswani, Shazeer, Parmar, Uszkoreit, Jones,
  Gomez, Kaiser, and Polosukhin}{Vaswani et~al\mbox{.}}{2017}]%
        {vaswani2017attention}
\bibfield{author}{\bibinfo{person}{Ashish Vaswani}, \bibinfo{person}{Noam
  Shazeer}, \bibinfo{person}{Niki Parmar}, \bibinfo{person}{Jakob Uszkoreit},
  \bibinfo{person}{Llion Jones}, \bibinfo{person}{Aidan~N Gomez},
  \bibinfo{person}{\L~ukasz Kaiser}, {and} \bibinfo{person}{Illia Polosukhin}.}
  \bibinfo{year}{2017}\natexlab{}.
\newblock \showarticletitle{Attention is All you Need}. In
  \bibinfo{booktitle}{\emph{Advances in Neural Information Processing Systems
  30}}. \bibinfo{pages}{5998--6008}.
\newblock


\bibitem[\protect\citeauthoryear{Wang, Wang, Joty, and Hoi}{Wang
  et~al\mbox{.}}{2021}]%
        {wang2021codet5}
\bibfield{author}{\bibinfo{person}{Yue Wang}, \bibinfo{person}{Weishi Wang},
  \bibinfo{person}{Shafiq Joty}, {and} \bibinfo{person}{Steven~CH Hoi}.}
  \bibinfo{year}{2021}\natexlab{}.
\newblock \showarticletitle{Codet5: Identifier-aware unified pre-trained
  encoder-decoder models for code understanding and generation}.
\newblock \bibinfo{journal}{\emph{arXiv preprint arXiv:2109.00859}}
  (\bibinfo{year}{2021}).
\newblock


\bibitem[\protect\citeauthoryear{Wang, Yao, Kwok, and Ni}{Wang
  et~al\mbox{.}}{2020}]%
        {wang2020generalizing}
\bibfield{author}{\bibinfo{person}{Yaqing Wang}, \bibinfo{person}{Quanming
  Yao}, \bibinfo{person}{James~T Kwok}, {and} \bibinfo{person}{Lionel~M Ni}.}
  \bibinfo{year}{2020}\natexlab{}.
\newblock \showarticletitle{Generalizing from a few examples: A survey on
  few-shot learning}.
\newblock \bibinfo{journal}{\emph{ACM computing surveys (csur)}}
  \bibinfo{volume}{53}, \bibinfo{number}{3} (\bibinfo{year}{2020}),
  \bibinfo{pages}{1--34}.
\newblock


\bibitem[\protect\citeauthoryear{White, Tufano, Vendome, and Poshyvanyk}{White
  et~al\mbox{.}}{2016}]%
        {white2016deep}
\bibfield{author}{\bibinfo{person}{Martin White}, \bibinfo{person}{Michele
  Tufano}, \bibinfo{person}{Christopher Vendome}, {and} \bibinfo{person}{Denys
  Poshyvanyk}.} \bibinfo{year}{2016}\natexlab{}.
\newblock \showarticletitle{Deep learning code fragments for code clone
  detection}. In \bibinfo{booktitle}{\emph{Proceedings of the 31st IEEE/ACM
  International Conference on Automated Software Engineering}}. ACM,
  \bibinfo{pages}{87--98}.
\newblock


\bibitem[\protect\citeauthoryear{Xian, Lampert, Schiele, and Akata}{Xian
  et~al\mbox{.}}{2018}]%
        {xian2018zero}
\bibfield{author}{\bibinfo{person}{Yongqin Xian}, \bibinfo{person}{Christoph~H
  Lampert}, \bibinfo{person}{Bernt Schiele}, {and} \bibinfo{person}{Zeynep
  Akata}.} \bibinfo{year}{2018}\natexlab{}.
\newblock \showarticletitle{Zero-shot learning—a comprehensive evaluation of
  the good, the bad and the ugly}.
\newblock \bibinfo{journal}{\emph{IEEE transactions on pattern analysis and
  machine intelligence}} \bibinfo{volume}{41}, \bibinfo{number}{9}
  (\bibinfo{year}{2018}), \bibinfo{pages}{2251--2265}.
\newblock


\bibitem[\protect\citeauthoryear{Zhang, Zhang, Shen, and Gu}{Zhang
  et~al\mbox{.}}{2022}]%
        {zhang2022diet}
\bibfield{author}{\bibinfo{person}{Zhaowei Zhang}, \bibinfo{person}{Hongyu
  Zhang}, \bibinfo{person}{Beijun Shen}, {and} \bibinfo{person}{Xiaodong Gu}.}
  \bibinfo{year}{2022}\natexlab{}.
\newblock \showarticletitle{Diet Code is Healthy: Simplifying Programs for
  Pre-Trained Models of Code}. In \bibinfo{booktitle}{\emph{Proceedings of the
  2022 The ACM Joint European Software Engineering Conference and Symposium on
  the Foundations of Software Engineering}} (Singapore, Singapore)
  \emph{(\bibinfo{series}{ESEC/FSE 2022})}.
\newblock


\end{thebibliography}

\newpage

\appendix
\setcounter{page}{1}
\onecolumn
\section{Further Examples of Code Transformations}
\lstset{escapechar=@,style=CustomJavaWoNumbers}
\begin{figure*}[!htb]
\centering
{
\begin{tabular}{p{0.46\linewidth}|p{0.46\linewidth}}
\hline
\multicolumn{2}{c}{Example : 1} \bigstrut\\
\hline
\begin{center}
    \footnotesize \underline{1. Input}
\end{center}
\begin{lstlisting}
int maxVal ( int value1 , int VAR_2 , int target ) { 
    if ( value1 > VAR_2 ) { 
        for ( int i = value1 ; i < value1 ; i ++ ) { 
            return value1 > target ? value1 : target; 
        } 
        return value1 ; 
    } 
    else 
        return VAR_2 > target ? VAR_2 : target ; 
}
\end{lstlisting}    &  
\begin{center}
    \footnotesize  \underline{2. PLBART output}
\end{center}
\begin{lstlisting}
maxVal ( int value1 , int VAR_2 , int target ) { 
    if ( value1 > VAR_2 ) { 
        for ( int i = value1 ; i < value1 ; i ++ ) { 
            return value1 > target ? value1 : target; 
        } 
        return value1 ; 
    } 
    else 
        return VAR_2 > target ? VAR_2 : target ; 
}
\end{lstlisting}\vspace{1mm}
\\
\hlineB{1}
\begin{center}
    \footnotesize  \underline{3. \tool output}
\end{center}
\begin{lstlisting}
int maxVal ( int value1 , int @\highlight{value2}@ , int target ) { 
    if ( value1 > @\highlight{value2}@ ) { 
        for ( int i = value1 ; i < value1 ; i ++ ) { 
            return value1 > target ? value1 : target; 
        } 
        return value1 ; 
    } 
    else return @\highlight{value2}@ > target ? @\highlight{value2}@ : target ; 
}
\end{lstlisting}    &  
\begin{center}
    \footnotesize  \underline{4. CodeT5 output}
\end{center}
\begin{lstlisting}
if ( value1 > VAR_2 @\errorcode{~~~~~~}@
    value1 = VAR_2 ;
    value1 = VAR_2 ;
    value1 ;
    }
    return value1 ;
@\errorcode{~~~\}~~\}}@
\end{lstlisting}\bigstrut[b]\\
\hline
\multicolumn{2}{c}{Example : 2} \bigstrut\\
\hline
\begin{center}
    \footnotesize \underline{1. Input}
\end{center}
\begin{lstlisting}
int maxVal ( int value1 , int value2 , int target ) { 
    if ( value1 > value ) { 
        for ( int i = value1 ; i < value1 ; i ++ ) { 
            return value1 > target ? value1 : target; 
        } 
        return value1 ; 
    } 
    else 
        return value2 > target ? value2 : target ; 
} 
\end{lstlisting}    &  
\begin{center}
    \footnotesize  \underline{2. PLBART output}
\end{center}
\begin{lstlisting}
maxVal ( int value1 , int value2 , int target ) { 
    if ( value1 > value ) { 
        for ( int i = value1 ; i < value1 ; i ++ ) { 
            return value1 > target ? value1 : target; 
        } 
        return value1 ; 
    } 
    else 
        return value2 > target ? value2 : target ; 
} 
\end{lstlisting}\vspace{1mm}
\\
\hlineB{1}
\begin{center}
    \footnotesize  \underline{3. \tool output}
\end{center}
\begin{lstlisting}
int maxVal  ( int value1 , int value2 , int target ) { 
    if ( value1 > value ) { 
        for ( int i = value1 ; i < value1 ; i ++ ) { 
            @\highlight{\textbf{if} ( value1 > target ) \{ }@
                @\highlight{\textbf{return} value1 ;}@ 
            @\highlight{\}}@
            @\highlight{\textbf{else} \{}@
                @\highlight{\textbf{return} target ;}@ 
            @\highlight{\}}@
            return value1 ; 
        }
    }
    else @\highlight{\textbf{if} ( value2 > target ) \{}@
        @\highlight{\textbf{return} value2 ; }@
    } 
    @\highlight{\textbf{else} \{}@
        @\highlight{\textbf{return} target ;}@
    @\highlight{\}}@
}
\end{lstlisting}    &  
\begin{center}
    \footnotesize  \underline{4. CodeT5 output}
\end{center}
\begin{lstlisting}
int int i = value2 ; i ++ ) {
    return value2 ; 
}
@\errorcode{\} \}}@
@\errorcode{value1 value2 ; value1 ; \} value2 ; \} \}}@
\end{lstlisting}\bigstrut[b]\\
\hline
\end{tabular}
}

 \caption{Example of input generated code by different pre-trained models.}
\label{fig:further_example}
\end{figure*}

\lstset{escapechar=@,style=CustomJavaWoNumbers}
\begin{figure*}[!htb]
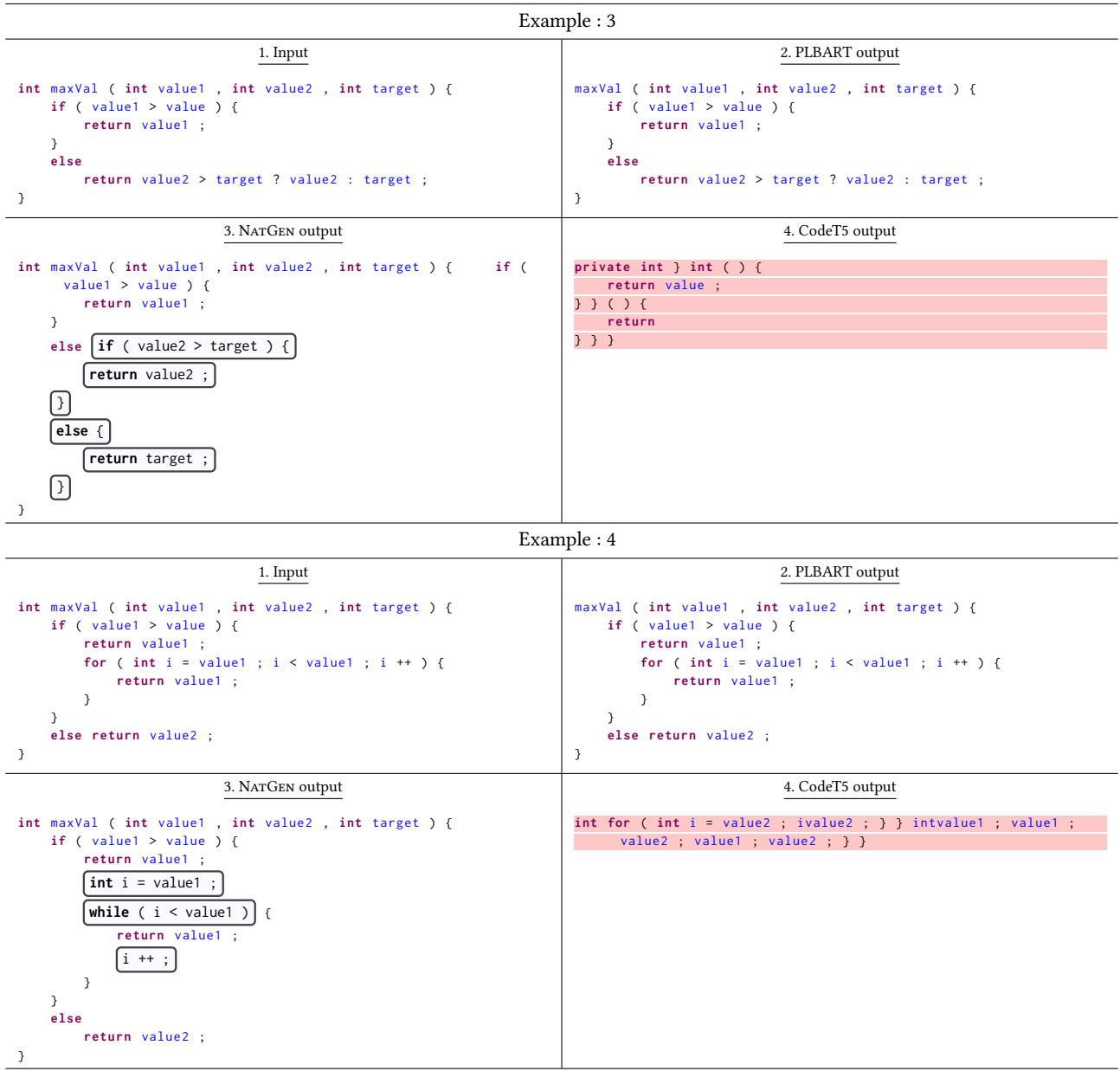

\centering
{
\begin{tabular}{p{0.46\linewidth}|p{0.46\linewidth}}
\hline
\multicolumn{2}{c}{Example : 3} \bigstrut\\
\hline
\begin{center}
    \footnotesize \underline{1. Input}
\end{center}
\begin{lstlisting}
int maxVal ( int value1 , int value2 , int target ) {
    if ( value1 > value ) { 
        return value1 ; 
    } 
    else 
        return value2 > target ? value2 : target ; 
}
\end{lstlisting}    &  
\begin{center}
    \footnotesize  \underline{2. PLBART output}
\end{center}
\begin{lstlisting}
maxVal ( int value1 , int value2 , int target ) {
    if ( value1 > value ) { 
        return value1 ; 
    } 
    else 
        return value2 > target ? value2 : target ; 
}
\end{lstlisting}\vspace{1mm}
\\
\hlineB{1}
\begin{center}
    \footnotesize  \underline{3. \tool output}
\end{center}
\begin{lstlisting}
int maxVal ( int value1 , int value2 , int target ) {     if ( value1 > value ) { 
        return value1 ; 
    } 
    else @\highlight{\textbf{if} ( value2 > target ) \{ }@
        @\highlight{\textbf{return} value2 ;}@ 
    @\highlight{\}}@
    @\highlight{\textbf{else} \{}@
        @\highlight{\textbf{return} target ;}@ 
    @\highlight{\}}@
}
\end{lstlisting}    &  
\begin{center}
    \footnotesize  \underline{4. CodeT5 output}
\end{center}
{
\lstset{escapechar=@,style=CustomJavaWoNumbers, backgroundcolor=\color{redish}}
\begin{lstlisting}
private int } int ( ) { 
    return value ; 
} } ( ) { 
    return 
} } }
\end{lstlisting}
}
\bigstrut[b]\\
\hline
\multicolumn{2}{c}{Example : 4} \bigstrut\\
\hline
\begin{center}
    \footnotesize \underline{1. Input}
\end{center}
\begin{lstlisting}
int maxVal ( int value1 , int value2 , int target ) {
    if ( value1 > value ) { 
        return value1 ; 
        for ( int i = value1 ; i < value1 ; i ++ ) { 
            return value1 ; 
        } 
    } 
    else return value2 ; 
}
\end{lstlisting}    &  
\begin{center}
    \footnotesize  \underline{2. PLBART output}
\end{center}
\begin{lstlisting}
maxVal ( int value1 , int value2 , int target ) {
    if ( value1 > value ) { 
        return value1 ; 
        for ( int i = value1 ; i < value1 ; i ++ ) { 
            return value1 ; 
        } 
    } 
    else return value2 ; 
}
\end{lstlisting}\vspace{1mm}
\\
\hlineB{1}
\begin{center}
    \footnotesize  \underline{3. \tool output}
\end{center}
\begin{lstlisting}
int maxVal ( int value1 , int value2 , int target ) {
    if ( value1 > value ) { 
        return value1 ; 
        @\highlight{\textbf{int} i = value1 ;}@
        @\highlight{\textbf{while} ( i < value1 )}@ {
            return value1 ;
            @\highlight{i ++ ; }@ 
        } 
    } 
    else 
        return value2 ; 
}
\end{lstlisting}    &  
\begin{center}
    \footnotesize  \underline{4. CodeT5 output}
\end{center}
{
\lstset{escapechar=@,style=CustomJavaWoNumbers, backgroundcolor=\color{redish}}
\begin{lstlisting}
int for ( int i = value2 ; ivalue2 ; } } intvalue1 ; value1 ; value2 ; value1 ; value2 ; } }

\end{lstlisting}
}\bigstrut[b]\\
\hline
\end{tabular}
}

 \caption{Example of input generated code by different pre-trained models (contd.)}
\label{fig:further_example_2}
\end{figure*}

\section{Pre-training progression for all metrics}


\begin{figure}[!htb]
    \centering
    \begin{subfigure}{0.25\linewidth}
    \centering
    \includegraphics[width=.95\linewidth]{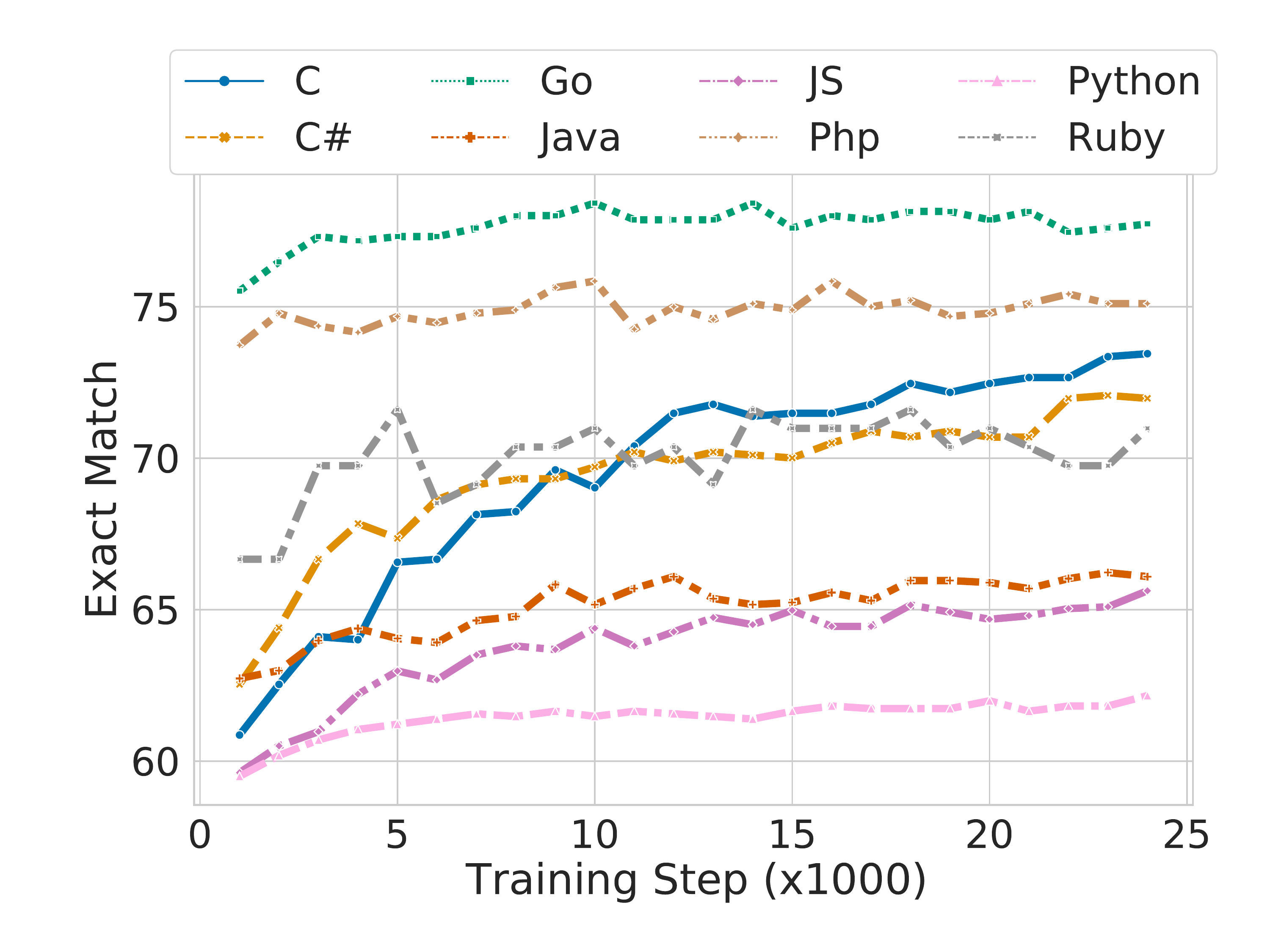}
    \label{fig:lang_to_accuracy_pretrain_perf}
    \end{subfigure}%
    \begin{subfigure}{0.25\linewidth}
    \centering
    \includegraphics[width=.95\linewidth]{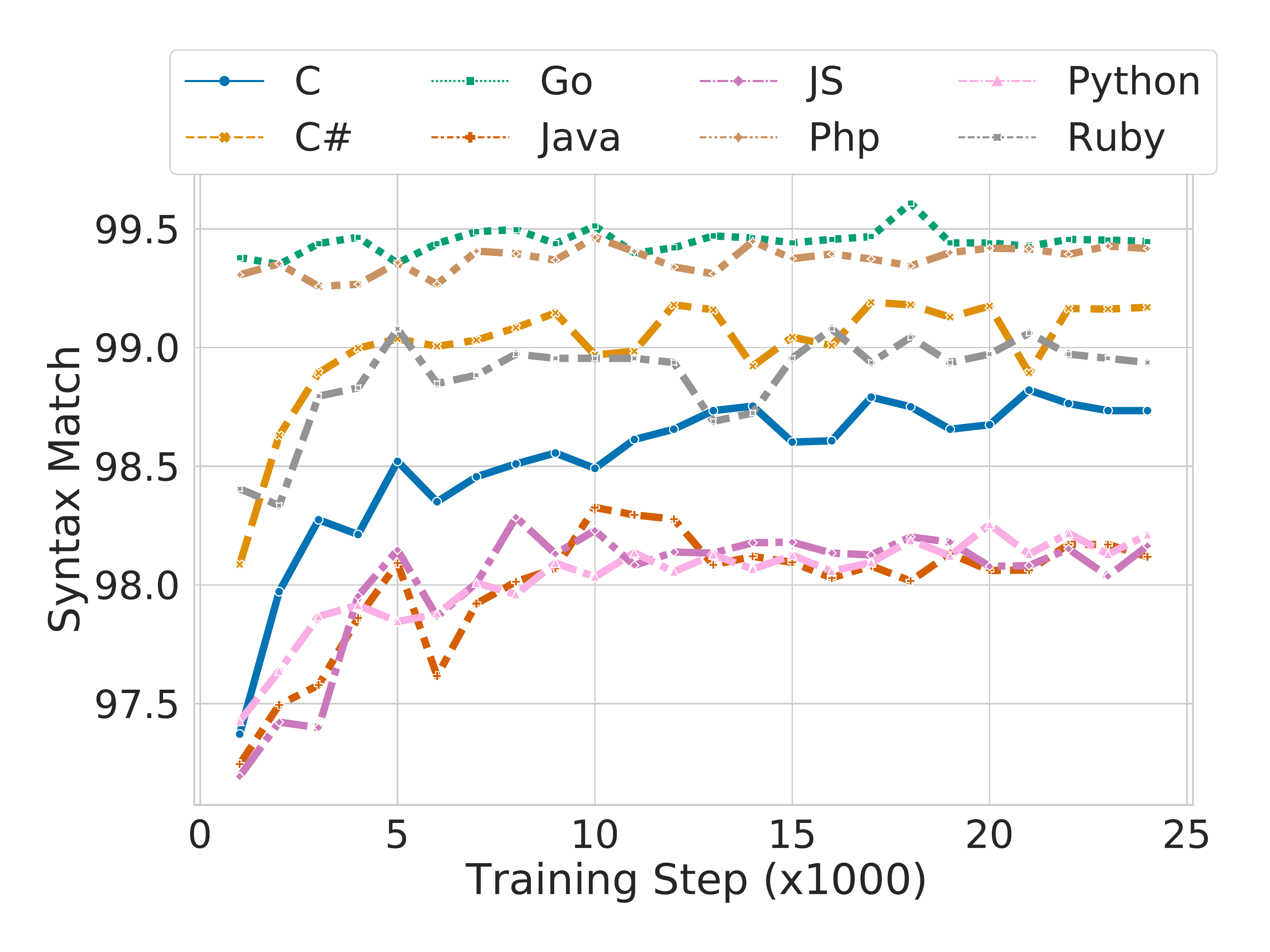}
    \label{fig:lang_to_syntax_pretrain_perf}
    \end{subfigure}%
    \begin{subfigure}{0.25\linewidth}
    \centering
    \includegraphics[width=.95\linewidth]{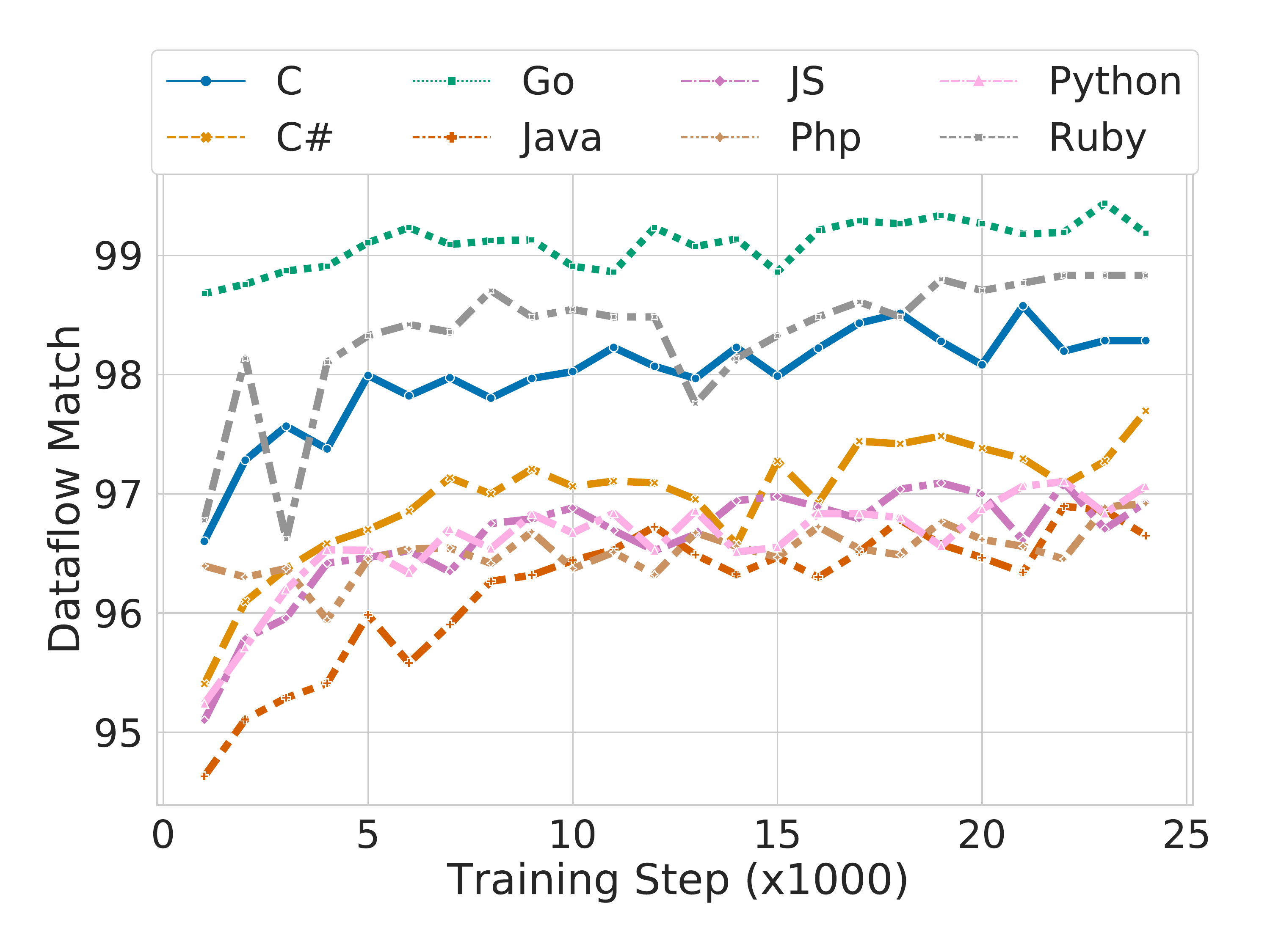}
    \label{fig:lang_to_dataflow_pretrain_perf}
    \end{subfigure}%
    \begin{subfigure}{0.25\linewidth}
    \centering
    \includegraphics[width=.95\linewidth]{images/rq1/lang_to_CodeBleu.pdf}
    \label{fig:lang_to_codebleu_pretrain_perf}
    \end{subfigure}
    \vspace{-5mm}
    \caption{Progression of Different metrics of different language in Validation dataset over number pre-training steps.}
    \label{fig:pretrain_all_over_step}
\end{figure}


\end{document}